\documentclass{aa}  
\usepackage{natbib,twoopt}
\usepackage{graphicx}	% Including figure files

\usepackage{savesym}
\usepackage{amsmath}
\savesymbol{iint}
\usepackage{txfonts}
\restoresymbol{TXF}{iint}
\usepackage{hyperref}
\newcommand{\dgr}{$^{\circ}$}
\newcommand{\co}{$^{12}$CO~}
\newcommand{\coo}{C$^{18}$O~}
\newcommand{\nthp}{N$_{2}$H$^{+}$~}

\newcommand{\kms}{km s$^{-1}$~}
\usepackage{comment}
\usepackage{lscape}
\bibpunct{(}{)}{;}{a}{}{,} %% natbib format for A&A and ApJ
\begin{document}
   \title{Distance, magnetic field and kinematics of a filamentary cloud LDN 1157}
   \subtitle{} 
   \author{Ekta Sharma \inst{1, 2}\thanks{E-mail: ektasharma.astro@gmail.com}
          \and
          G. Maheswar\inst{1}
          \and
          A. Soam\inst{3}
          \and
          Chang Won Lee\inst{4,5}
          \and
          Kim, Shinyoung\inst{4}
          \and
          Tuhin Ghosh\inst{6}
          \and
          Tej, A\inst{7}
          \and
          Kim, G\inst{9}
          \and
          S. Neha\inst{8} 
          \and Piyali Saha\inst{1}}

   \institute{Indian Institute of Astrophysics, Koramangala, Bangalore 560034, India
         	 \and
             Department of Physics and Astrophysics, University of Delhi, Delhi 110007, India
             \and
             SOFIA Science Centre, USRA, NASA Ames Research Centre, MS-12, N232, Moffett Field, CA 94035, USA
             \and
             Korea Astronomy \& Space Science Institute, 776 Daedeokdae-ro, Yuseong-gu, Daejeon, Republic of Korea
             \and
             University of Science and Technology, Korea, 217 Gajeong-ro, Yuseong-gu, Daejeon 34113, Republic of Korea
             \and
             School of Physical Sciences, National Institute of Science Education and Research, HBNI, Jatni 752050, Odisha, India 
             \and
             Indian Institute of Space Science and Technology, Valiamala P.O., Thiruvananthapuram - 695547 Kerala, India
             \and
             School of Space Research, Kyung-Hee University, 1732, Deogyeong-daero, Giheung-gu, Yongin-si, Gyeonggi-do 17104, Korea
             \and
             Nobeyama Radio Observatory, National Astronomical Observatory of Japan, National Institutes of Natural Sciences, Nobeyama, Minamimaki, Minamisaku, Nagano 384-1305, Japan
             }

   \date{Received......, Accepted.....}
% \abstract{}{}{}{}{} 
% 5 {} token are mandatory
  \abstract
   {LDN 1157 is one of the several clouds situated in the cloud complex, LDN 1147/1158. The cloud presents a coma-shaped morphology with a well-collimated bipolar outflow emanating from a Class 0 protostar, LDN 1157-mm, residing deep inside the cloud. }
   {The main goals of this work are (a) to map the inter-cloud magnetic field (ICMF) geometry of the region surrounding LDN 1157 to investigate its relationship with the cloud morphology, with the outflow direction and with the core magnetic field (CMF) geometry inferred from the mm- and sub-mm polarization results from the literature, and (b) to investigate the kinematic structure of the cloud.}
   {We carried out optical (R-band) polarization observations of the stars projected on the cloud to map the parsec-scale magnetic field geometry and made spectroscopic observations of the entire cloud in $^{12}$CO, C$^{18}$O and N$_{2}$H$^{+}$ (J=1-0) lines to investigate its kinematic structure.}
   {We obtained a distance of 340$\pm$3 pc to the LDN 1147/1158, complex based on the \textit{Gaia} DR2 parallaxes and proper motion values of the three young stellar objects (YSOs) associated with the complex. A single filament of $\sim1.2$ pc in length (traced by the \textit{Filfinder} algorithm) and $\sim0.09$ pc in width (estimated using the \textit{Radfil} algorithm) is found to run all along the coma-shaped cloud. Based on the relationships between the ICMF, CMF,  filament orientations, outflow direction, and the presence of an hour-glass morphology of the magnetic field, it is likely that the magnetic field had played an important role in the star formation process in LDN 1157. LDN 1157-mm is embedded in one of the two high density peaks detected using the \textit{Clumpfind} algorithm. Both the detected clumps are lying on the filament and show a blue-red asymmetry in the \co line. The \coo emission is well correlated with the filament and presents a coherent structure in velocity space. Combining the proper motions of the YSOs and the radial velocity of LDN 1147/1158 and an another complex LDN 1172/1174 which is situated $\sim2$\dgr~east of it, we found that both the complexes are moving collectively toward the Galactic plane. The filamentary morphology of the east-west segment of LDN 1157 may have formed as a result of mass lost by ablation due to the interaction of the moving cloud with the ambient interstellar medium.}
   {}
   \keywords{ISM: clouds; polarization: dust; ISM: magnetic fields; ISM: individual objects: L1157
              }
     \maketitle
% 
%-------------------------------------------------------------------
%%%%%%%%%%%%%%%%%%%%%%%%%%%%%%%%%%%%%%%%%%%%%%%%%%

%%%%%%%%%%%%%%%%%%%%%%%%%%%%%%%%%%%%%%%%%%%%%%%%%%

\section{Introduction} \label{sec:intro}
Observations from the \textit{Herschel} Space Observatory, with its unprecedented sensitivity and resolution, revealed the omnipresence of a deep network of filaments \citep{2010A&A...518L.102A, 2010A&A...518L.103M} both in quiescent and in star forming molecular clouds \citep[e.g., ][]{2009ApJ...700.1609M, 2010A&A...518L.102A, 2011A&A...529L...6A, 2013MNRAS.432.1424K, 2016A&A...591A..90R}. These filaments are often found to extend out from dense star-forming hubs \citep{2009ApJ...700.1609M, 2012ApJ...756...10L, 2013ApJ...779..121G} with more star forming sites distributed along their length \citep[e.g., ][]{2009ApJ...700.1609M, 2012A&A...540L..11S, 2014prpl.conf...27A, 2015A&A...574A.104T}. These filamentary structures are described as either stagnant gas produced due to large-scale compression flows \citep{2012A&A...541A..63P} or isothermal self-gravitating cylinders in pressure equilibrium with external medium \citep{2012A&A...542A..77F, 2013ApJ...776...62H} or are supported by helical magnetic fields \citep{2000ApJ...534..291F}. To gain more insight into the processes involved in the formation of these filaments, it is important to study their internal structure, magnetic field geometry and kinematics of gas.

Several models and simulations have been carried out to understand the processes such as iso-thermal-driven turbulence \citep{2002ApJ...576..870P, 2005A&A...436..585D, 2015ApJ...807...67M}, thermal \citep{2003ApJ...585L.131V, 2005A&A...433....1A, 2008ApJ...683..786H} or gravitational \citep{1996ApJ...467..321N, 1999PASJ...51..625U, 2014ApJ...789...37V} instabilities that could potentially influence the formation of filamentary structures in molecular clouds. Presence of filaments in unbound and non-star forming regions \citep{2010A&A...518L.104M} suggests a paradigm in which they represent the early stages of core and star formation. In this paradigm, the dense material within molecular clouds is first accumulated into dense filaments which then fragment to form star forming cores \citep[][]{2001MNRAS.327..715B, 2014ApJ...791..124G}. These fragments may continue to amass material through gravitational inflow resulting in a flow pattern parallel to the filament axis \citep[e.g., ][]{2001MNRAS.327..715B, 2006MNRAS.373.1091B}. 
Correlated magnetic fields on scales of 1-10 pc have been observed in interstellar clouds \citep[e.g., ][]{1976AJ.....81..958V, 1987ApJ...321..855H, 1986ApJ...309..619M, 1992ApJ...399..108G, 2004ApJ...603..584P, 2011ApJ...741...21C, 2015A&A...573A..34S, 2017MNRAS.464.2403S, 2017ApJ...849..157W, 2018ApJ...867...79C}. The magnetic fields may play an important role in regulating these flows. But how exactly the magnetic fields, turbulence and self-gravity compete or collaborate to form, first, filaments, then cores and finally stars is still unclear. 

According to \citet{2014ApJ...791..124G} and \citet{2016MNRAS.455.3640S}, filaments arise as a result of anisotropic global collapse of the clouds. Initially gas is accreted onto it from the surrounding environment expecting the magnetic field to be perpendicular to the filament axis. While the low density material is found to be aligned with the field lines, the dense filaments, however,  are seen perpendicular to the magnetic fields \citep{2016A&A...586A.135P, 2016A&A...586A.138P}. Such a trend is also seen in the polarization observations of background stars in optical \citep[e.g., ][]{2000AJ....119..923H, 2004ApJ...603..584P, 2008A&A...486L..13A} and near infrared \citep[e.g., ][]{1992ApJ...399..108G, 2011ApJ...741...21C} wavelengths. However, as the gas density increases, the gas tends to flow along the filaments dragging the field lines along with it thus making it parallel to the filament axis \citep{2018MNRAS.480.2939G}. Therefore, the relative orientation of filaments with respect to the local magnetic fields \citep[e.g., ][]{2009ApJ...704..891L}, kinematics of the gas both perpendicular and parallel to the filaments are some of the important observational signatures that are key to understand the manner in which the material was accumulated during the formation of filaments. Identifying and characterizing isolated filaments that are at their earliest evolutionary stage of star formation are crucial in this regard.

In this paper, we present results of a study conducted on an isolated star forming molecular cloud LDN 1157 (L1157). The cloud is located in the Cepheus flare region and is spatially \citep{1962ApJS....7....1L, 2002A&A...383..631D} and kinematically \citep{1997ApJS..110...21Y} associated with a complex containing a number of clouds namely, LDN 1147/1148/1152/1155/1157/1158 (hereafter L1147/1158 complex). There is an ambiguity in the adopted distance of the cloud. The most widely quoted distance of L1147/1158 is 325$\pm$13 pc \citep{1992BaltA...1..149S}. However, in several studies, 250 pc \citep[e.g., ][]{2007ApJ...670L.131L, 2016A&A...593L...4P} and 440 pc \citep[e.g., ][]{1996A&A...307..891G, 1996ApJ...463..642A} distances are also used. The L1157 harbors a cold, extremely red object, IRAS 20386+6751 (hereafter L1157-mm), classified as a Class 0 source \citep{1993ApJ...406..122A, 1996MmSAI..67..901A} having a bolometric luminosity of 11 $L_{\odot}$ and bolometric temperature between 60-70 K \citep{1997A&A...323..943G}. The source shows a well collimated bipolar outflow of $\sim5\arcmin$ spatial size \citep{2001A&A...372..899B} and a $\sim2\arcmin$ flattened envelope perpendicular to it \citep{2007ApJ...670L.131L}. The position angle of the outflow measured counterclockwise from the north is 161$^{\circ}$ \citep{2001A&A...372..899B} with an inclination angle of $\sim$10\dgr~\citep{1996A&A...307..891G}. The magnetic field orientation inferred from 1.3 mm polarization measurements shows an hourglass morphology with the central vectors showing a position angle of $\sim148^{\circ}$ measured counterclockwise from north \citep{2013ApJ...769L..15S}.

%*******************************************************************
\begin{figure}
\includegraphics[width=8.9cm,height= 8.6cm]{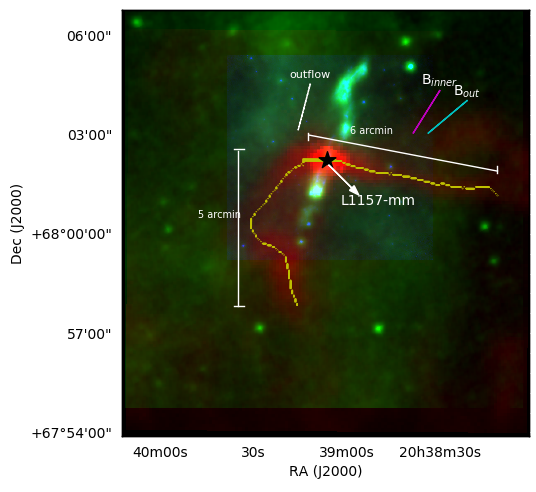}
\caption{Color composite image of the filamentary cloud L1157 made using \textit{Herschel} 250 $\mu$m (red), WISE 12 $\mu$m (green) and Spitzer 8 $\mu$m emission (blue). A filament structure in yellow color based on the dust column density (N(H$_{2}$)) distribution extracted using the \textit{Filfinder} algorithm is also shown. White segment shows orientation of the outflow, magenta and cyan segments represent the orientation of inner (traced for submm polarization emission measurements) and outer magnetic fields (traced for the optical polarization measurements of background stars), respectively.}\label{fig:rgb}
\end{figure}
%*******************************************************************

Using the column density map produced from the \textit{Herschel} images, we traced a single filament which runs almost in the east-west direction and then changes its direction towards the south. The filament is traced using the \textit{Filfinder} algorithm (described in section \ref{sec:fil}). Both the east-west and the north-south segments of the filament are found to be $\sim5\arcmin$ in length. In Fig. \ref{fig:rgb}, we show a color composite image of the region containing the cloud L1157 produced using the \textit{Herschel} 250 $\mu$m (red), \textit{WISE} 12 $\mu$m (green) and \textit{Spitzer} 8 $\mu$m (blue) emission. Emission from the protostar (identified using black star symbol), the bipolar outflow originating from it and a well defined filament structure extending to the west of the protostar is conspicuous in Fig. \ref{fig:rgb}. The age estimates of L1157-mm, $\sim150$ kyr \citep{2001A&A...372..899B, 2005ApJS..156..169F, 2008ApJ...681L..21A}, suggest that the star formation has just recently got initiated in L1157 and hence the conditions that led this cloud to form star(s) may still be preserved. Additionally, absence of any active high mass star formation in the vicinity of L1157 \citep{2009ApJS..185..451K} presents a simple case of isolated low mass star formation occurring in a quiescent environment.  

In this work, we first estimated the distance to L1147/1158 complex using the recently released \textit{Gaia} DR2 parallax and proper motion values of the young stellar object (YSO) candidates associated with it. To investigate the role played by the parsec-scale magnetic field in the formation of L1157, we made optical R-band polarization measurements of stars that are projected on a region of $20\arcmin\times20\arcmin$ field that includes the cloud. Due to the lack of kinematic information in the \textit{Herschel} continuum dust maps, we made molecular line observations of the region containing the filament structure in L1157 in $^{12}$CO, C$^{18}$O, N$_{2}$H$^{+}$ ($\rm J=1-0$) lines. Finally, using the \textit{Gaia} DR2 proper motion values of the YSOs associated with the L1147/1158 and L1172/1174 complexes (an another complex situated 2\dgr~east of L1147/1158) and using the radial velocities of the cloud, we determined the motion of the complexes in space and discuss a possible origin of the filament in L1157. The paper is organized in the following manner. We begin with a brief description of the observations, data and the reduction procedure in \S \ref{sec:obs_dr}. The results from the polarization and molecular line observations are presented in \S \ref{sec:results}. Discussion on the results obtained is presented in \S \ref{sec:discussion}. We concluded the paper with a summary of the results obtained in this work in \S  \ref{sec:summary}.

%############################OBSERVATIONS AND DATA REDUCTION
\section{Observations and Data Reduction}\label{sec:obs_dr}
\subsection{Optical Polarimetry}
The polarimetric observations presented here were carried out over several nights in November 2015 at the Cassegrain focus of the 104-cm Sampurnanand Telescope, Nainital, India. We used the Aries IMaging POLarimeter (AIMPOL) which incorporates an achromatic rotatable half wave plate (HWP) as modulator and a Wollaston prism beam splitter as the analyzer. The fast axis of the HWP and the axis of the Wollaston prism are kept perpendicular to the optical axis of the system. The fast axis of the HWP is rotated at four different angles 0\degr, 22.5\degr, 45\degr and 67.5\degr. This provides two images of the object on the field of CCD camera which is of TK 1024$\times$1024 pixel$^2$ size \citep[see ][]{2004BASI...32..159R}. The plate scale and the gain of the CCD used are 1.48$\arcsec$ per pixel and  11.98 e$^{-}$ per ADU (Analog to digital unit) respectively. The noise created while CCD is read out is 7.0 e$^{-1}$. 

Standard Johnson R$_{kc}$ filter with $\lambda$ $_{eff}$ as 0.76 $\mu$m was used for polarimetric observations. The spatial resolution (FWHM) corresponds to 2-3 pixels on CCD. The data reductions are carried out using software developed in Python to identify the ordinary and corresponding extraordinary images of each object in the field of view. The photometry is carried out using aperture photometry provided by the \emph{Image Reduction \& Analysis Facility (IRAF)} package. The intensities of ordinary and extraordinary images in the observed field are extracted to calculate the required ratio R($\alpha$). This ratio is defined as:

\begin{equation}
R(\alpha) = \frac{\frac{I_{e}(\alpha)}{I_{o}(\alpha)}-1}{\frac{I_{e}(\alpha)}{I_{o}(\alpha)}+1} = P \cos(2\theta - 4\alpha)
\end{equation}  
where P is the strength of the total linearly polarized light, $\theta$ is the polarization angle in the plane of sky and $\alpha$ is the angle of half-wave plate fast axis with respect to the axis of the Wollaston prism. The P and $\theta$ are calculated using normalized stokes parameters q1, u1 and q2, u2 at angles 0\degr, 22.5\degr, 45\degr, 67.5\degr respectively.

Standard polarized and unpolarized stars selected from \cite{1992AJ....104.1563S} are observed routinely to correct the polarization position angle offset and the instrumental polarization respectively. The instrumental polarization derived from the unpolarized standards is found to be $\sim0.1\%$ \citep[e.g., ][]{2008MNRAS.388..105M, 2013MNRAS.432.1502S}. The polarization angles of the standard stars obtained from our observations were compared with those of the standard star values given in the \cite{1992AJ....104.1563S} and the difference was applied as a correction to the polarization angles. In Table \ref{tab:pol_log} we show the log of the polarization observations.

%*********************************************************************************************
\begin{table}  %% observation log for standard stars
\caption{Polarized standard stars observed in$\ R_{\mathrm{kc}}$ band.}\label{tab:pol_log}
\begin{tabular}{c c c} \hline
 Date of Observation & P $\pm$ $\epsilon_{\mathrm{P}}$  & $\theta$ $\pm$ ${\epsilon_{\theta}}$ \\ 
          &            $(\%)$                &              (\degr)             \\ \hline

\multicolumn{3}{c}{\bf HD236633} ($^{\dagger}$Standard values: 5.38 $\pm$ 0.02$\%$, 93.04 $\pm$ 0.15\degr) \\
 3 November 2015 & 4.9 $\pm$ 0.1 &  99$\pm$1 \\
 15 November 2015 & 5.0 $\pm$ 0.1 &  101$\pm$ 1\\
 16 November 2015 & 5.1 $\pm$ 0.1 &  101$\pm$ 1 \\
 17 November 2015 & 4.9 $\pm$ 0.1 &  101$\pm$ 1 \\ \\

\multicolumn{3}{c} {\bf BD+59\degr389} ($^{\dagger}$Standard values: 6.43 $\pm$ 0.02$\%$, 98.14 $\pm$ 0.10\degr) \\
  2 November 2015 & 6.3 $\pm$ 0.1 &  105$\pm$ 1 \\
  3 November 2015 & 6.2 $\pm$ 0.1 &  105$\pm$ 1 \\
  15 November 2015 & 6.0 $\pm$ 0.1 &  106$\pm$1 \\
  16 November 2015 & 5.9 $\pm$ 0.1 &  105$\pm$1 \\
  17 November 2015 & 6.2 $\pm$ 0.1 &  106$\pm$1\\ \hline
 \end{tabular}
 
$^{\dagger}$Values in the R$-$band taken from \cite{1992AJ....104.1563S}.
 \end{table}
 
%###############################################################################################3

\subsection{Radio Observations}
The molecular chemistry is different at various layers of the molecular cloud which makes it difficult to discuss the  kinematics with a single molecular tracer. We have chosen a set of molecules $^{12}$CO ($\rm J=1-0$), C$^{18}$O ($\rm J=1-0$), N$_{2}$H$^{+}$ ($\rm J=1-0$) to be observed with the same telescope to constitute a homogeneous set of same calibration. Because of the high difference in dipole moment of N$_{2}$H$^{+}$ and $^{12}$CO molecules, this data is sensitive to high as well as low density regions by using these tracers.
The filamentary structure of L1157 cloud has been mapped with these tracers using 13.7 m diameter single dish radio facility at Taeudek Radio Astronomy Observatory (TRAO) which is located at Korea Astronomy and Space Science Institute (KASI) in Daejeon, South Korea. It operates in the wavelength range of 85-115 GHz.
Observations were taken using the new receiver system SEQUOIA-TRAO (Second QUabbin Observatory Imaging Array-TRAO). It consists of high-performing 16 pixel MMIC preamplifiers in a 4$\times$4 array. The pointing accuracy was achieved to be $\leq$ $5^{''}$ by using standard X cygnus source in SiO line. The position switch mode is employed to subtract the sky signals. At 115 GHz the beam size (HPBW) of the telescope is about $45^{''}$ and the fraction of the beam pattern subtending main beam (beam efficiency) is 51$\pm$2\% . The system temperature was 550 K-600 K during the observations. 
The back-end system with fast fourier transform spectrometer has 4096 $\times$ 2 channels at 15 kHz resolution ($\sim$0.05 \kms at 110 GHz). As the optical system provides 2-side-band, two different lines can be observed simultaneously. \coo line which reveals the dynamics of high density regions of cloud was simultaneously observed with \co. The observations were performed using the On-The-Fly (OTF) mapping technique, covering a region of $12^{'}$ $\times$ $12^{'}$ for $^{12}$CO, C$^{18}$O and 8$^{'}$ $\times$ 8$^{'}$ region for N$_{2}$H$^{+}$ in $\rm J=1-0$ transition. The center of the maps was 20h39m06.19s +68$\degr$02$\arcmin$15.09$\arcsec$.
The signal-to-noise ratio (SNR) for \co line at the position 20h39m12.837s +68$\degr$01$\arcmin$06$\arcsec$ is found to be $\sim18$ with the peak ${T_{A}}^{\ast}$ as 3.87 K and rms as 0.22 K at velocity resolution 0.06 \kms.
The spectra were reduced using CLASS software of the IRAM GILDAS software package. A first order polynomial was applied to correct the baseline in the final spectra. The resulting 1$\sigma$ rms noise levels in ${T_{A}}^{\ast}$ scale are $\sim$ 0.3 K for \co (1-0) and 0.1 K  for \coo line, respectively. The final data cubes have cell size of 22$\arcsec$ and 0.06 \kms velocity channel width. 

%%%%##############################  Results #####################################

\section{RESULTS} \label{sec:results}

\subsection{Polarization results}
\begin{figure}
\includegraphics[width=8.3cm, height = 7.8cm]{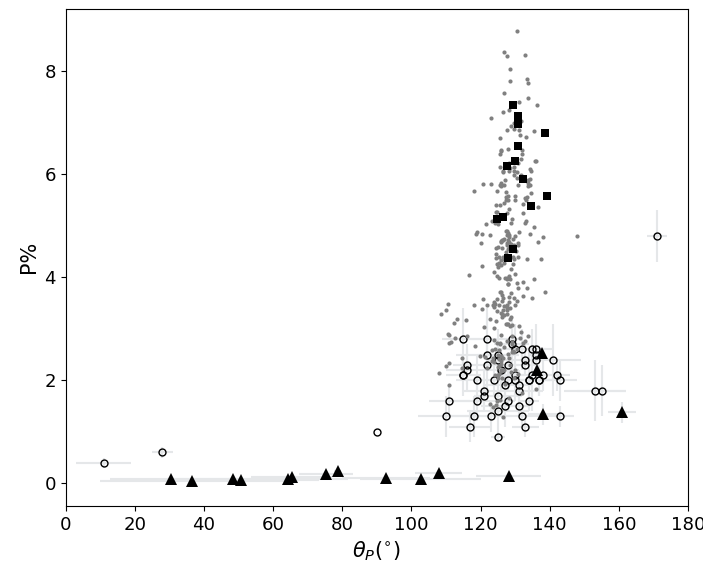}
\caption{The P$\%$ vs. $\theta_{P}$ for the 62 sources (open circles) projected over 0.3\degr $\times$ 0.3\degr area around the protostar, L1157-mm. The measurements are made in R$_{kc}$ filter. The \textit{Planck} polarization results (see \S\ref{sec:mag_field}) from within 1$\degr$ region around L1157 cloud are shown using filled circles in gray. The \textit{Planck} results from the region where we carried out the optical polarization observations are shown using the squares in black. Also shown are the polarization values (filled triangles) of the sources distributed in a circular region of 5\degr~radius about the protostar obtained from the \citet{2000AJ....119..923H} catalog.}\label{fig:pol_pa}
\end{figure}

We made optical polarization measurements of 62 stars that are projected on an area of 0.3\degr $\times$ 0.3\degr around the protostar, L1157-mm. We show the measured degree of polarization (P\%) and polarization position angles ($\theta_{P}$) in Fig. \ref{fig:pol_pa} using open circles. The polarization measurements for which the ratio of P\% and its corresponding error is greater than 3 are plotted. For the majority of the sources, P\% ranges between $\sim1-2$\% and the $\theta_{P}$ ranges between $\sim110^{\circ}-140^{\circ}$. The mean values of P\% and $\theta_{P}$ of the sources showing P\%$\geqslant1$ are 2.1 and 129$^{\circ}$, respectively and the corresponding standard deviation values are 0.6 and 11$^{\circ}$, respectively. Also shown are the polarization values of 16 sources selected from a circular area of 5\degr~radius about the protostar. Though there are 18 sources present within our search radius, we rejected two sources, namely, HD 200775 and HD 193533. The HD 200775 is an intermediate mass Herbig Be star causing a reflection nebulosity, NGC 7023. The HD 200775 is situated at a distance of 357$\pm$6 pc \citep{2018AJ....156...58B} and the P\% and $\theta_{P}$ values given in the \citet{2000AJ....119..923H} catalog is $\sim0.8\pm$0.2 and $\sim92^{\circ}\pm7^{\circ}$. It is highly likely that the polarization measurements are affected by the presence of the intense emission from the nebulosity surrounding the star due to scattering. The second star, HD 193533 is an M3III and classified as a variable star in the Simbad database. The distance, P\% and $\theta_{P}$ values for this star are 301$\pm$5 pc, 0.3$\pm$0.05 and 142$^{\circ}\pm4^{\circ}$, respectively. 

%********************************************************************
%********************************************************************

\subsection{Identification of Filaments and Clumps}\label{sec:fil}
The whole L1147/1158 complex was observed by the \textit{Herschel} telescope. The PACS \citep{2010A&A...518L...2P} and SPIRE \citep{2010A&A...518L...3G} instruments were used to observe the region simultaneously at 70 $\mu$m, 160 $\mu$m, 250 $\mu$m, 350 $\mu$m and 500 $\mu$m wavelengths as a part of the Gould Belt Survey \citep{2010A&A...518L.102A}. The 160-500 $\mu$m \textit{Herschel} images were used to construct an H$_{2}$ column density map and a dust temperature map of the entire complex at the spatial resolution of 36$^{''}$. The units of SPIRE images which is in MJysr$^{-1}$ were changed to Jy pixel$^{-1}$ using \textit{convertImageunit} task in Herschel Interactive Processing Environment (HIPE). For the SED fitting on pixel-to-pixel basis, all the maps (PACS 160 $\mu$m, SPIRE 250 $\mu$m, 350 $\mu$m, 500  $\mu$m) were convolved to the 500 $\mu$m image using the kernels from \cite{2011PASP..123.1218A} and regridded to a pixel scale of 14$^{''}$. The background flux density (I$_{bg}$) was determined using values from the pixels in a relatively darker patch in the sky. The emission of every pixel is assumed to be represented by a modified blackbody emission: 
\begin{equation}
S_{\nu} (\nu) = \Omega (1-\exp(-\tau_{\nu})) B_{\nu}(\nu,T_{d}),
\end{equation} with
\begin{equation}
\tau_{\nu}= 0.1\mu m_{H} \kappa_{\nu} N_{H_{2}},
\end{equation}
where S$_\nu$($\nu$) is the observed flux density for a given frequency $\nu$ and solid angle $\Omega$, $\tau$($\nu$) is the optical depth, B($\nu$,T$_{d}$) is the Planck function, T$_{d}$ is the dust temperature, m$_{H}$ is mass of hydrogen, $\mu_{m_{H}}$ is the mean molecular weight taken as 2.8 \citep{2008A&A...487..993K},  N$_{H_{2}}$ is the column density for hydrogen. All the fluxes have been normalized to Jy pixel$^{-1}$. For opacity, we assumed a functional form of $\kappa$$_{\nu}$ = 0.1 ($\frac{\nu}{1000GHz})$$^{\beta}$ where $\beta$ is the spectral emissivity index and the value is taken as 2 \citep{2010ApJ...708..127S}. The derived column density and dust temperature maps were regridded using Astronomical Image Processing System (AIPS) to 3\arcsec. Because we are modelling the cold dust emission longward of 160 $\mu$m, the fit was relatively poorer near the protostar where contribution from warm dust would also be present. Hence, a single blackbody model cannot be used.

\begin{figure}
   \includegraphics[width=8.3cm, height=7cm]{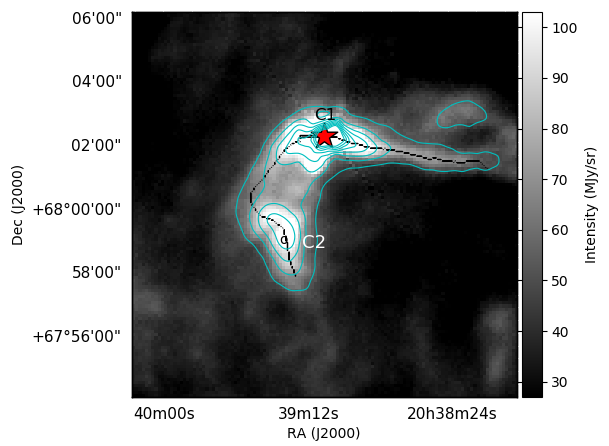}
  \caption{Contours of column density, N($H_{2}$), in cyan color overlayed on the \textit{Herschel} 250 $\mu$m gray scale emission. The star symbol in red and the small circle in black identify the position of the clump C1 which harbors the protostar L1157-mm and the clump C2, respectively. The contours are shown from the levels of 3-20$\sigma$ ($\sigma$ $\approx$ 7$\times$ 10$^{20}$ cm$^{-2}$). }\label{fig:250-HGBS}
\end{figure}

To characterize the filament properties, we used \textit{FilFinder} algorithm to extract the filaments from the column density map. The \textit{FilFinder} algorithm was developed to extract filamentary structures in clouds observed by the \textit{Herschel}  \citep{2010A&A...518L.102A}. The extraction is performed by reducing the regions of interest to topological skeletons based on specified threshold intensities. Each element of the skeletons therefore represents the medial position within the extents of the required region \citep{2015MNRAS.452.3435K}. The emission structures in L1157 was flattened to a percentile of 99 to smooth the bright features in the image. While creating masks, the global threshold was taken as $\sim2.1\times10^{21}$ cm$^{-2}$ (3$\sigma$ above the background, $\sigma\sim$ 7.0$\times$ 10$^{20}$ cm$^{-2}$) with size threshold as 300 pix$^{2}$. The masks were reduced to skeletons using medial axis transform which extracts one single filament. A single filament of $\sim$1.2 pc in length that runs all along the coma-structure of the cloud is traced.  For the purpose of analysis, we divided the filament into east-west and north-south segments. The orientation of the east-west segment is found to be 77\degr and a curvature of 76\degr with respect to the north increasing eastward. The extracted filament is shown in Fig. \ref{fig:rgb} and Fig \ref{fig:250-HGBS}.  

%*********************************************************************
%*********************************************************************
\begin{figure*}
\centering
\includegraphics[width=16cm, height = 8.4cm]{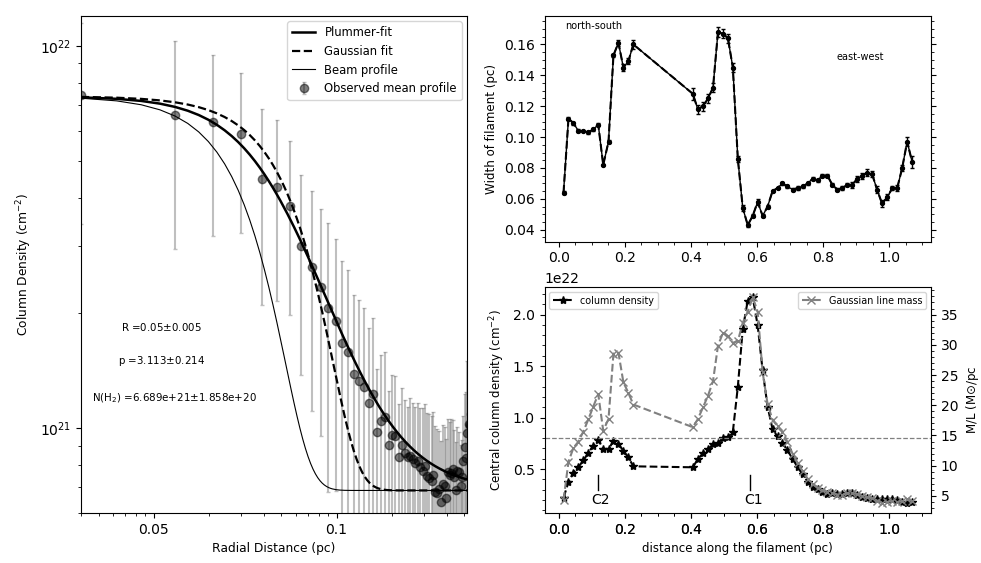}
\caption{{\bf Left:} Mean radial column density profile of the L1157 filament (in gray points) measured perpendicular to the crest of filament shown in fig \ref{fig:rgb}. The gray error bars mark the $\pm1\sigma$ dispersion of the distribution of radial profiles along the spine of filament. The solid black curve shows the best-fit plummer model fitted on the mean radial profile. The black dashed curve marks the best-fit Gaussian function to the inner radius of the profile. The thin solid black curve represents the gaussian profile of the beam. {\bf Right upper panel:} Deconvolved Gaussian FWHM of the L1157 filament as a function of position along the crest of filament (starting from southern core towards central protostar). {\bf Right lower panel:} The central column density along the crest of the filament obtained from the best-fit Plummer model in black dashed line. Background-subtracted mass-per-unit length calculated from the Gaussian fit (in gray dashed line). The gray dashed horizontal line indicates the critical mass-per-unit length or line mass of an isothermal filament in equilibrium as 2c$_{s}^{2}$/G $\sim$ 15 M$_{\odot}$ pc$^{-1}$ at 10 K.}\label{fig:radfil-sub}
\end{figure*}
%*********************************************************************

We used the well-known \textit{Clumpfind} \citep{1994ApJ...428..693W} routine to identify high density regions on the filament from the column-density map. Based on the \textit{Clumpfind} routine, we obtained two clumps: C1 and C2 which are lying on the filament (Fig. \ref{fig:250-HGBS}). The centers of these clumps are 20h39m06.79s +68$\degr$02$\arcmin$12.27$\arcsec$ and 20h39m20.349s +67$\degr$59$\arcmin$03.74$\arcsec$ respectively with a typical uncertainty in the positions of $\sim10\arcsec$ \citep{2017A&A...606A.102F}. While the clump C1, located on the east-west segment where the protostar L1157-mm is currently forming, the clump C2 is found to be located on the north-south segment of the filament. Based on the absence of any 70 $\mu$m source associated with the C2, we classify it as starless. We noted that if we use 250 $\mu$m emission map instead of the column density map, the \textit{Clumpfind} algorithm resolves C2 into two separate clumps. This may be due to the high spatial resolution of the 250 $\mu$m emission map compared to the column density map. Because our molecular line observations are also finding a single density peak at the position of C2 due to relatively coarse spatial resolution, we considered C2 as a single clump in our study. 

Radial profiles and widths of the filaments are two of the most important properties of prime interest to understand the dominant physics (gravity, turbulence, magnetic field orientation) involved in their formation. We constructed the column density profiles of the filament identified in L1157 using the publicly available package, \textit{Radfil} \citep{2018ApJ...864..152Z}. The derived filament mask and the spine of the column density map derived from \textit{Filfinder} were supplied as an input. The spine was smoothed to get a continuous distribution in column density. The crest of the filament was sampled at an interval of 40 pixels (0.18 pc) which corresponds to three times the beam width (0.061 pc at 340 pc). Therefore, the mean profile was constructed by averaging the profiles of the perpendicular cuts made at nine positions along the filament and setting the \textit{Fold = True} in the \textit{Radfil} to add all the profiles towards positive radial distance. We fixed the fit distance from 0.0 to 0.5 pc and evaluated the background at a distance of 0.5-0.6 pc from the filament crest (out of all possible trials conducted using the \textit{Radfil}, the minimum value of the background column density is estimated from this distance range). A zeroth-order polynomial fit was applied to the background subtraction before making the fits to the profile. The diameter of the flat inner plateau is found to be 2R$_{flat}$ = 0.126$\pm$0.003 pc. The observed mean column density profile was fit by a Gaussian model over the inner radius of 0.05 pc. The power-law index of the best-fit Plummer model is p = 3.1$\pm$0.2 while the mean deconvolved width of the best-fit Gaussian model is FWHM = 0.09 pc. 

An elongated structure showing minimum values of both aspect ratio and column density contrast with respect to the background value is normally identified as a filament \citep{2019A&A...621A..42A}. The aspect ratio, defined as $l_{fil}/W_{fil}$, of the sole filament identified in L1157 is 1.2/0.09$\approx14$. The intrinsic column density contrast, N$^{o}_{H_{2}}$/N$^{bg}_{H_{2}}$, of the filament is estimated to be $\approx8$, where N$^{o}_{H_{2}}$ (= N$^{fil}_{H_{2}}$-N$^{bg}_{H_{2}}$), is the column density amplitude of the filament. The column densities of the pixels that form the extracted filament structure are averaged to get a representative value for the whole filament.
One of the important consequences of obtaining column density profile is to calculate the mass per unit length. We derived the mass per unit length (M$_{line}$) for each position along the filament using the best fit Gaussian parameters: central column density (N$_{H_{2}}$) and standard deviation ($\sigma$). The right lower panel of Fig. \ref{fig:radfil-sub} shows the distribution of the background subtracted Gaussian mass-per-unit length along the crest of filament at every cut which was taken at an interval of 3 pixels ($\sim$ 0.015 pc). The dashed horizontal line marks the critical mass-per-unit length which characterises an isothermal cylindrical filament in equilibrium. The extreme end of the north-south segment of the filament (20h39m17s +67d57m56s) was taken as the starting point of the filament. The right upper panel in Fig. \ref{fig:radfil-sub} shows the deconvolved FWHM derived from the Gaussian fitting of the mean column density radial profile as a function of distance along the filament crest. Although the cloud reflects as a single filament derived from the \textit{Filfinder} algorithm, differences are found in the inner widths of the north-south and the east-west branches. The characteristic width of the east-west branch is more well-defined and constrained as compared to that of the north-south branch. The positions where we could not fit the radial profile with a well-defined Gaussian function are not included in the plot. The small dip near the C2 corresponds to the region between the two peaks seen in 250 $\mu$m \textit{Herschel} map (FWHM $\sim18\arcsec$) but barely noticeable in the column density map (FWHM $\sim36\arcsec$).

\subsection{Molecular line analysis}
The CO isotopologues are commonly used to probe the gas at different densities. While the most abundant isotopologue, $^{12}$CO is considered to trace the most diffuse and external gas of molecular  clouds, its rarer counterpart, \coo ($\rm J=1-0$) line having a critical density of 2.4 $\times$ 10$^4$ cm$^{-3}$, is one of the best tracers of high column- and volume densities without getting saturated. However, \coo is found to disappear from the gas phase in high, chemically evolved regions due to its condensation onto the surface of the dust grains \citep{1999ApJ...523L.165C, 2002ApJ...570L.101B, 2017ApJ...849...80C}. The \nthp molecular line, on the other hand, is considered to be the most efficient tracer of dense cores in clouds because the abundance of this molecule gets enhanced when CO condenses onto the dust grains \citep{2007ARA&A..45..339B}. 

The kinematic information from the observed molecular lines is extracted by fitting Gaussian profiles to all our spectra using programs developed in Python. We examined individual spectrum and fitted independently using one single component at each individual position in \coo lines. We obtained a total of 1089 spectra in $^{12}$CO, \coo and 306 spectra in \nthp lines from the field containing the cloud L1157.

Various fundamental properties of the cloud like excitation temperature, optical depth, and the number density are derived using spectral data obtained at different positions of the cloud. Two assumptions have been made to estimate the column density (N$_{18}$) based on C$^{18}$O and $^{12}$CO lines; the molecules along the line of sight possess a uniform excitation temperature for $\rm J=1-0$ transition and the $\rm J=1-0$ excitation temperatures of the two isotopic species are equal \citep[e.g., ][]{1978ApJS...37..407D, 1994ApJ...435..279S}. The emission from $^{12}$CO molecule is optically thick and common excitation temperature (T$_{ex}$) is calculated from the peak $^{12}$CO brightness temperature using the expression, $T_b = T_R^{12}/ \eta_{eff}$, where $\eta_{eff}$ is the beam efficiency of TRAO telescope. The brightness temperature, $T_b^{12}$, is calculated as follows, 
\begin{equation}
T_b^{12}= T_{0}^{12} [f_{12}(T_{ex}) - f_{12}(2.7)],
\end{equation}
where T$_{0}^{12}$ is the temperature corresponding to the energy difference between the two levels given by T$_{0}^{12}$ = h$\nu_{12}$/k, here $\nu_{12}$ is the frequency for $\rm J=1-0$ transition. In the same manner, the optical depth of the C$^{18}$O line ($\tau_0^{18}$) is also calculated from the peak brightness temperature and the excitation temperature (T$_{ex}$) using the expression,
\begin{equation}
T_R^{18} = T_0^{18} [ f_{18}(T_{ex}) - f_{18}(2.7)] [1- \exp(-\tau_0^{18})].
\end{equation}
The column density along the line of sight is calculated as follows, %using the equation
\begin{equation}
N_{18}= \frac{3h\triangle  v_{18}}{8\pi^3\mu ^2} {\frac{\tau ^{18}Q}{[1-\exp{-\frac{T_0}{T_{ex}}]}}},
\end{equation}
where $\rm \Delta$v$_{18}$  is the line full width at half maxima in velocity units, $\mu$ is the permanent dipole moment of the molecule, h is the Planck constant and Q is the partition function. Again an assumption has been made regarding the partition function, which depends upon the excitation temperatures of all significantly populated states of the molecule like:
\begin{equation}
Q= 1+ \sum_{L=1}^{+\infty} (2L+1)  \prod_{J=0}^{L} \exp[-h\nu(J)/kT_{ex}(J)],
\end{equation}
where $\nu$(J) and T$_{ex}$(J) are the frequency and excitation temperature of the transition $\rm J=1-0$, respectively. So we assume that all the levels have same T$_{ex}$. The partition function can be written as 
Q=  $\frac{2 T_{ex} ^{18}}{T_0 ^{18}}$.
But all the lines are not fitted to a perfect Gaussian. Thus the chi-square minimization has been done to determine the goodness-of-fit. We obtained integrated intensity (moment 0) maps of $^{12}$CO, C$^{18}$O using spectral data cubes with position-position velocity information. The values below 3$\sigma$ were not considered to obtain summation over the channels. The rms value of each line map was calculated as $\sim\sqrt{N}\delta$vT$_{rms}$, where $N$ is the number of channels used for integrating the emission, $\delta$v is the velocity resolution and T$_{rms}$ is the noise of the line profile.

%***********************************************
\begin{figure*}
\centering
  \includegraphics[width=\textwidth,height=12.9cm]{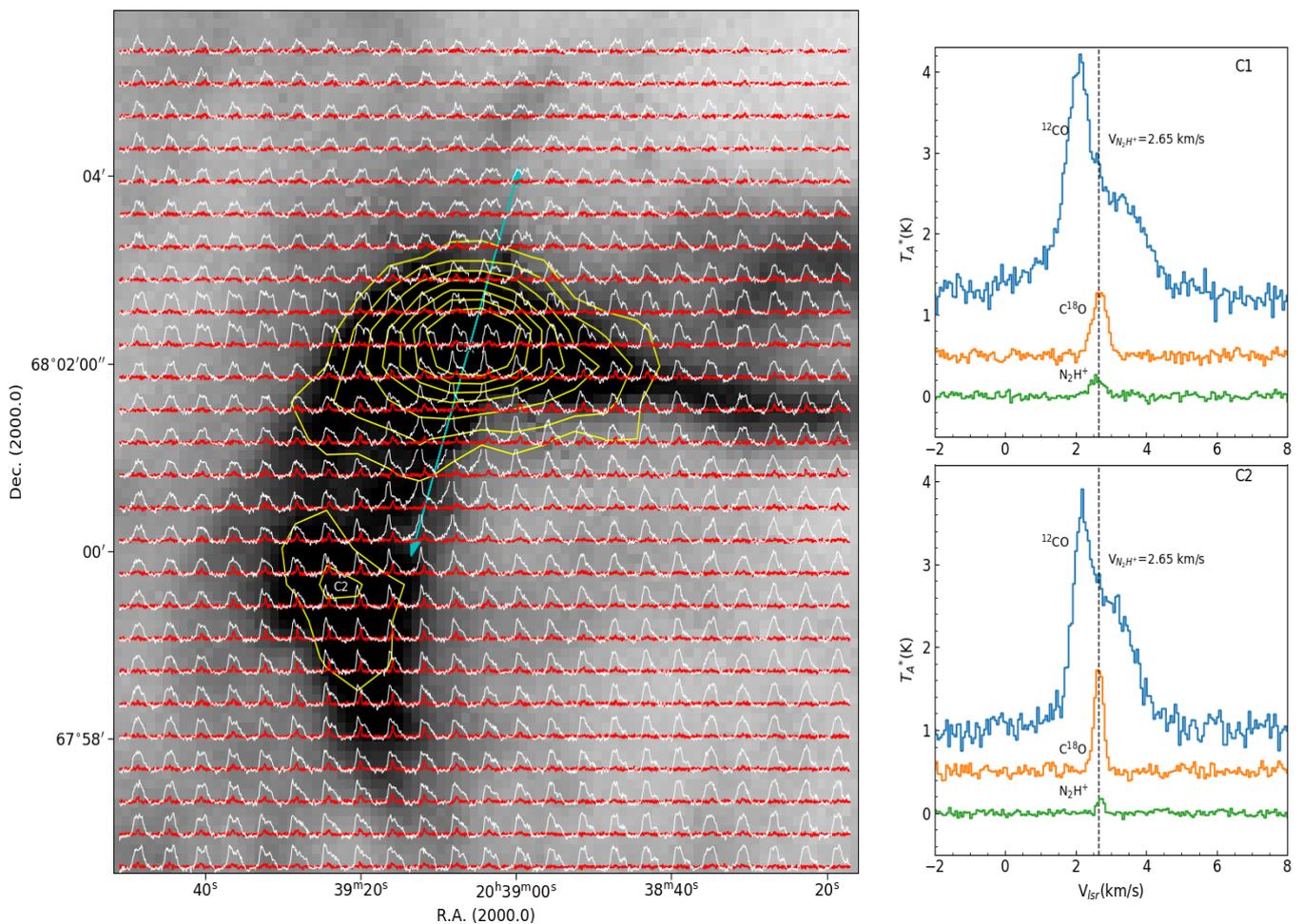}
  \caption{The distribution of \co (white) and \coo (red) profiles over the 9\arcmin $\times$ 9\arcmin region. On the left-hand panel, the background image is \textit{Herschel} 250 $\mu$m emission for L1157 cloud where contours of N$_{2}$H$^{+}$ (1-0) line overlayed in yellow color. The position of core C1 and clump C2 are marked in white color. The extent of outflow is marked by the arrow in cyan. Contour levels start from 4$\sigma$ in steps of 3$\sigma$ where $\sigma\sim$ 0.05 K \kms. The small windows on the right-hand panel shows the average spectra of \co, \coo, \nthp (1-0) (isolated component) lines for C1 (top) and C2 (bottom). The average was taken over the half-maximum contour of the intensity map of \nthp emission for C1 and C2. The dashed line indicates the velocity of \nthp obtained from the Gaussian hyperfine fitting of its seven hyperfine components. }\label{fig:profiles-full}
\end{figure*}
%***********************************************

The characteristic shape of the observed line profiles is a critical factor in determining the physical state of molecular gas in a region. Fig. \ref{fig:profiles-full} on left-hand panel shows $^{12}$CO (white) and \coo (1-0) (red) line profiles plotted together at different positions on the cloud and on the right-hand panel, the average \co, \coo and \nthp (isolated component) profiles over the half-maximum contour of the intensity map of \nthp emission. The \co emission is detected from all over the observed region and conspicuously, shows diverse line profiles with asymmetric structure detected in a large area. The $^{12}$CO lines show wide line-width and two velocity components especially on the cloud. This could be either due to the self-absorption by the optically thick material at the systematic velocity of the cloud, if there is only a single cloud component, or due to the presence of additional velocity components along the line-of-sight. The significant emission of \coo line ($\geq$3$\sigma$) is obtained towards the high column density region seen in the \textit{Herschel} dust map as shown in Fig. \ref{fig:250-HGBS}. The \coo line profiles show an optically thin feature i.e., a single Gaussian component. The absorption of double peaked profiles of $^{12}$CO lines coincides with the single peak of \coo emission confirming the existence of only a single cloud component along the line of sight.

The \coo line width is found to be much narrower than that of $^{12}$CO. The presence of well studied outflow \citep{1992ApJ...392L..83U} is evident as broad high velocity wings on both sides of the \co line profiles when compared visually with the shape of \coo Gaussian component. Interestingly many positions along the filament and surrounding regions around C1 and C2, the profiles show blue-red asymmetry (section \ref{sec:fil_gas}). The \nthp emission is detected towards both C1 and C2 (contours shown in yellow color).

\section{Discussion}\label{sec:discussion}

\subsection{Distance of L1147/1158 complex}\label{sec:dist}

One of the most direct methods of finding distances to molecular clouds is to estimate distances of the YSOs that are associated with the cloud \citep[e.g., ][]{2007ApJ...671..546L, 2018ApJ...869L..33O}. The stellar parallax measurements obtained for these YSOs from the \textit{Gaia} DR2 \citep{2018A&A...616A...2L} offer an unprecedented opportunity to estimate distances to molecular clouds with improved accuracy and precision. A total of 6 YSO candidates have been identified so far toward the direction of L1147/1158 \citep{1998ApJS..115...59K, 2009ApJS..185..198K}. We found a \textit{Gaia} counterpart for 3 of the 6 YSO candidates well within a search radius of 1$\arcsec$. The \textit{Gaia} results are presented in Table \ref{tab:yso_dist_pm}. We obtained their distances from the \citet{2018AJ....156...58B} catalog and proper motion values in right ascension ($\mu_{\alpha\star}=\mu_{\alpha}$ cos$\delta$) and in declination ($\mu_{\delta}$) from the \citet{2018A&A...616A...1G} catalog. Within the L1147/1158 complex \citep{1997ApJS..110...21Y}, the dark cloud, L1155, harbors two YSOs, namely, 2MASSJ20361165+6757093 and IRAS 20359+6745 and in cloud L1158, at its north-east edge, an another YSO candidate, PV Cephei, is located. Presence of a bright nebulosity associated with IRAS 20359+6745 \citep{2003A&A...399..141M} and the PV Cephei \citep{1991MNRAS.249..131S} are a clear evidence of their association with their respective clouds. No detection was found in the \textit{Gaia} DR2 database for the other three YSO candidates, namely L1148-IRS, which is associated with L1148 \citep{1998ApJS..115...59K}, IRAS 20353+6742 which is associated with L1152 \citep{1988BAAS...20..693B} and L1157-mm in L1157. 

The mean value of the distance calculated from the three YSOs is 340 pc with a dispersion of 3 pc. The mean (standard deviation) values of the $\mu_{\alpha\star}$ and $\mu_{\delta}$ for them are 7.806 (0.326) mas/yr and -1.653 (0.229) mas/yr respectively. Similar distance and proper motion values shown by all the three YSOs indicate that they are spatially and kinematically associated with each other and that the complex is also located at a distance of $340\pm3$ pc from us. Similar values of V$_{lsr}$ ($\approx2.6$ \kms) shared by the individual clouds of the complex \citep{1991A&A...249..493H, 1997ApJS..110...21Y, 2014ApJ...788..108S} also supports the above argument. \citet{1992BaltA...1..149S}, based on Vilnius photometry which gives two dimensional classification and extinction suffered by stars, estimated distances to the L1147/1158 cloud complex. Using 10 reddened stars in the direction of L1147/1158 they estimated a distance of 325$\pm$13 pc to the cloud. 

The degree of P\% measured in the optical wavelengths made using the pencil-beam of a starlight passing through the interstellar medium is often found to correlate with the extinction ($A_{V}$) measured to that line of sight upto at least an $A_{V}$ of $\sim3$ magnitudes \citep{1989AJ.....98..611G, 1999A&A...349..912H}. Therefore as the distance to the observed stars increases, the column of the dust grains present along the pencil-beam also increases leading to a gradual increase in the P\% provided no significant depolarization is occurring along the path. When the starlight passes through a molecular cloud, the measured P\% will have an additional contribution from the dust grains present in it. This will lead to a sudden increase in the values of P\% for the stars background to the cloud while the foreground stars will show P\% due to the contribution from the foreground ISM alone. Therefore, the presence of a molecular cloud can be inferred from the measured polarization of the foreground and the background stars \citep[e.g., ][]{1987Ap&SS.133..355C, 1989AJ.....98..611G, 1993MNRAS.265....1A, 1998MNRAS.300..497R, 2007A&A...470..597A, 2016A&A...588A..45N}. 

%********************************************************************
\begin{figure}
\includegraphics[width=8.3cm, height = 12cm]{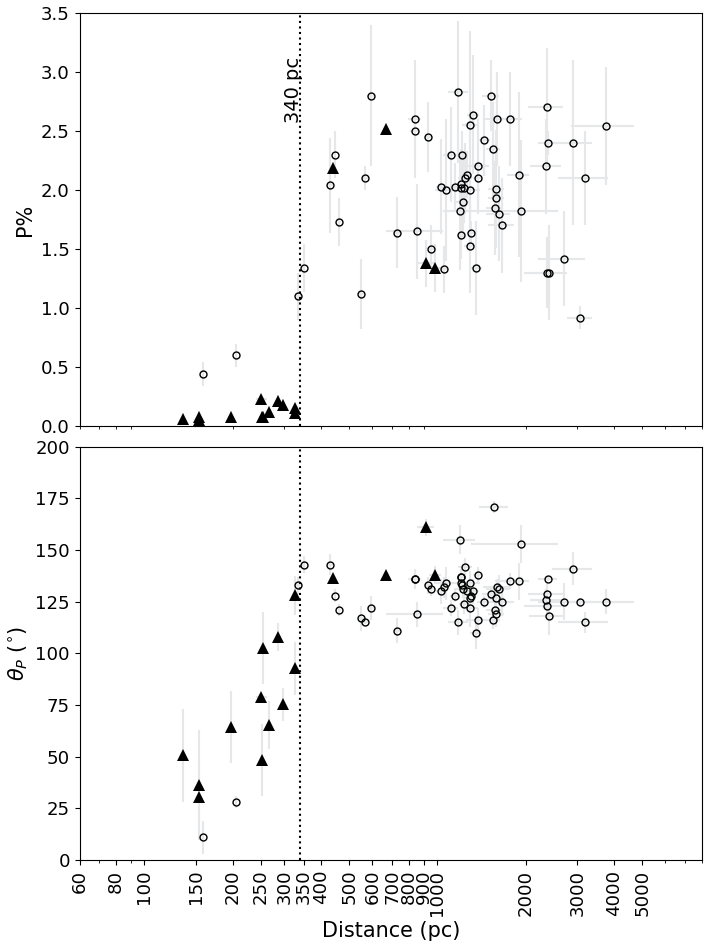}
\caption{{\bf Upper panel:} The P$\%$ vs. distance of the sources for which we made polarization measurements (open circles in black). The distances are obtained from \citet{2018AJ....156...58B} catalog. Polarization measurements of the field stars (filled triangles) are obtained from the \citet{2000AJ....119..923H} catalog. The vertical line is drawn at 340 pc. {\bf Lower panel:} The variation of polarization position angles of stars as a function of their distances. }\label{fig:pol_dist}
\end{figure}
%******************************************************************** 

In Fig. \ref{fig:pol_dist} we show P\% for the 62 stars, observed by us, as a function of their distances that are obtained from the \citet{2018AJ....156...58B} catalog by making a search for the \textit{Gaia} counterparts within a search radius of 1$\arcsec$. For all the sources, we found a counterpart well within  1$\arcsec$ from our input coordinates. The sources selected from the \citet{2000AJ....119..923H} catalog are shown using black triangles. The distances for these stars are also obtained from the \citet{2018AJ....156...58B} catalog. Up to a distance of $\sim340$ pc, the P\% of sources obtained from the \citet{2000AJ....119..923H} catalog show very low values. The weighted mean values of P\% and $\theta_{P}$ for the sources located at distances $\leqslant340$ pc are 0.1$\pm$0.05 and 65$^{\circ}\pm$29$^{\circ}$, respectively. For the four sources located beyond 340 pc, the weighted mean values of P\% and $\theta_{P}$ are found to be 1.6$\pm$0.4 and 148$^{\circ}\pm$11$^{\circ}$, respectively. The $\theta_{P}$ of the sources from the \citet{2000AJ....119..923H} catalog are found to show a systematic change from $\sim25^{\circ}$ to $\sim125^{\circ}$ with the distance till about 340 pc. Beyond this distance, sources are found to show $\theta_{P}$ similar to those obtained for our target sources. Among the sources observed by us, only two sources are having distances less that 340 pc. The weighted mean value of P\% and $\theta_{P}$ of these two sources are 0.5$\pm$0.1 and 16$^{\circ}\pm8^{\circ}$, respectively. For the sources observed by us and located at or beyond 340 pc, the weighted mean value of P\% and $\theta_{P}$ are found to be 2.1$\pm$0.6 and 129$^{\circ}\pm11^{\circ}$, respectively. The distribution of P\% and $\theta_{P}$ as seen in Fig. \ref{fig:pol_dist} further supports the 340 pc distance estimated for the L1147/1158 cloud complex. 

\subsection{Magnetic field geometry in L1157}\label{sec:mag_field}

\begin{figure*}
\centering
\includegraphics[width=17cm, height=8cm]{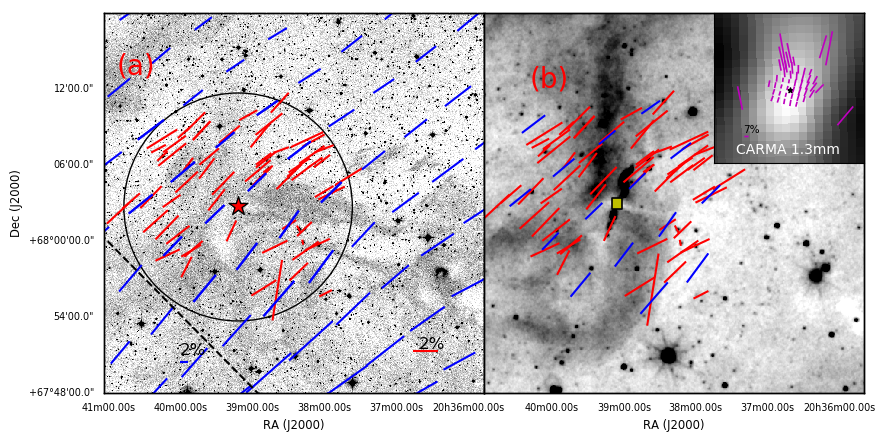}
\caption{\textbf{(a):} Optical polarization vectors (in red color) over plotted on image of 0.5$^{\circ}$ $\times$ 0.5$^{\circ}$ DSS image. The dashed line shows the direction of galactic plane. The circle shows the region of optical polarization observations. $\textit{Planck}$ polarization vectors are shown in blue color (inside and outside the circle). \textbf{(b):} \textit{WISE} 12 $\mu$m image for the same region in inverted scale. Optical (in red) and Planck (in blue) polarization vectors are overlayed. The box in yellow color around the location of protostar marks the region of submillimetre polarization observations observed in wavelength 1.3 mm using CARMA \citep{2013ApJ...768..159H} and the vectors are shown in inset (upper-right) by magenta color. The location of protostar in the inset is identified by black star symbol.}\label{fig:dss_her_pol_C18O}
\end{figure*}

The dynamics of interstellar dust grains can be affected by the presence of magnetic field. It was shown that a rotating non-spherical dust grains would tend to align with their long axis perpendicular to the interstellar magnetic field \citep{1951ApJ...114..206D, 1967ApJ...147..943J, 1979ApJ...231..404P, 1995ApJ...453..229L, 2012ARA&A..50...29C}. When an unpolarized starlight from a background star passed through regions of such aligned dust grains, they polarize the starlight by selectively absorbing the component parallel to the long axis of the grains. Thus the polarization position angles provide a sense of the plane-of-sky component of the magnetic field. As it is evident that most of the stars observed by us towards L1157 are background, the measured values of $\theta_{P}$ represent the magnetic field geometry of the cloud. The two stars in our sample which are foreground show higher P\% values when compared to those from the \citet{2000AJ....119..923H} catalog. We subtracted the mean P\% and $\theta_{p}$ values of these two foreground sources from the rest of the sources vectorially and found the values to be 2.3$\pm$0.8 and 127$^{\circ}\pm12^{\circ}$. Therefore, 127$^{\circ}\pm12^{\circ}$ is taken as the orientation of magnetic field in L1157 inferred through optical polarization. The magnetic field orientations thus obtained are overplotted (using vectors in red) on the DSS image as shown in Fig. \ref{fig:dss_her_pol_C18O} (a). The position of the protostar L1157-mm is shown using a star symbol in red. The length of the vectors corresponds to the P\% values and the orientations correspond to the $\theta_{P}$ values measured from the north increasing towards the east. The polarization measurements using background starlight in the optical wavelengths typically works only for the regions of low A$_{V}$ ($\lesssim5$). This is because, the dust grains present deep inside the molecular clouds though are efficient in diminishing background starlight, are believed to be not much efficient in polarizing the light in optical wavelengths \citep{1995ApJ...448..748G, 2012ARA&A..50...29C}. Therefore the optical polarization vectors shown in the Fig. \ref{fig:dss_her_pol_C18O} (a) basically trace the orientation of the magnetic fields at the envelope (low-density regions) of L1157. The magnetic field orientation found to be well ordered at $\sim0.2-2$ pc scales which suggests that the inter cloud magnetic field (ICMF) might have played an important role at least in the initial building up of the cloud. 

%**********************************************************************

%**********************************************************************
The magnetic field orientation of a region can also be inferred through the observations of polarized thermal emission from the dust grains \citep{1984ApJ...284L..51H, 1995ASPC...73...45G, 1999A&A...344..668G}. Far-infrared and submillimeter polarimetric observations made by the \textit{Planck} have been used not only to infer the direction of the Galactic magnetic field but also to put new constraints on the properties of dust  \citep{2015A&A...576A.106P, 2016A&A...586A.135P, 2016A&A...586A.138P}. \footnote{The entire sky was surveyed by the \textit{Planck} in nine frequency bands, from 30 GHz to 857 GHz with unprecedented sensitivity and angular resolutions varying from 30$\arcmin$ at 30 GHz to 4.8$\arcmin$ at 857 GHz. Of the nine, seven bands were sensitive to the polarized thermal emission from  the Galactic  dust.} We used the 353 GHz (850 $\mu$m) data, which is the highest-frequency polarization-sensitive channel of the \textit{Planck}, to produce the structure of the Galactic magnetic field in the vicinity of L1157. We selected a $\sim1\degr$ diameter image centered around the cloud and smoothed it down to the 8$\arcmin$ resolution to obtain good SNR. The results are shown in Fig. \ref{fig:pol_pa} using dots in gray. The P values range from $\sim$2\%-8\% with a mean value of 4.4\% and a standard deviation of 1.6\%. The dust emission is linearly polarized with the electric vector normal to the sky-projected magnetic field, therefore the polarization position angles are rotated by 90$\degr$ to infer the projected magnetic field.
The polarization position angles show a highly regular distribution with a mean value of 127$\degr$ and a standard deviation of 6$\degr$. The polarization vectors (magenta) are presented on the 0.5$^{\circ}$ $\times$ 0.5$^{\circ}$ DSS image as shown in Fig. \ref{fig:dss_her_pol_C18O} (a). The vectors in blue are those from a 0.3$^{\circ}$ $\times$ 0.3$^{\circ}$ circular region similar to where we carried out optical polarization observations. These results are shown using filled squares in black in Fig. \ref{fig:pol_pa}. It is noted that the positions showing relatively higher values of P show relatively less dispersion in $\theta_{P}$. The mean and the standard deviation of the source with P$\geqslant$4\% are 129$\degr$ and 4$\degr$ respectively. The projected magnetic field geometry inferred from the optical and the \textit{Planck} polarimetry are in good agreement with each other. Such an agreement between the magnetic field directions inferred from the optical and the \textit{Planck} are seen towards a number of clouds belonging to the Gould Belt \citep{2016A&A...596A..93S, 2019ApJ...871L..15G}.

\subsection{Magnetic field strength}
We used Davis-Chandrasekhar \& Fermi (DCF) \citep{1951PhRv...81..890D,1953ApJ...118..113C} method to estimate the plane-of-the-sky magnetic field strength of L1157. The DCF method is formulated as:
\begin{equation}
B_{pos}= 9.3 \times \sqrt{\frac{n_{H_{2}}}{cm^{-3}}} \times  \frac{\rm \Delta v}{km s^{-1}} \times \Big(\frac{\Delta  \phi }{1\degr}\Big)^{-1},
\end{equation}
where $\rm\Delta$v is FWHM of CO(1-0) line which is measured as $\sim$1.8 \kms. We used only those lines to measure FWHM which can be fitted with a single Gaussian function. Here n$_{H_{2}}$ is the volume density of L1157 which is found to be 1000 cm$^{-3}$. For calculating the volume density, we used column density of 8.0 $\times$ 10$^{20}$ cm$^{-2}$ and depth of the cloud as 0.3 pc. We considered the extent of column of the cloud material lying along the line-of-sight as three times the width of the filament which is around 0.1 pc. In the equation, $\rm\Delta\phi$ is dispersion in the distribution of polarization position angles which is measured as 11$\degr$. The magnetic field strength in the envelope of L1157 is found to be $\sim$ 50 $\mu$G. By propagating the uncertainties in measured position angle and velocity dispersion values, we calculated the uncertainty in magnetic field strength as $\approx$0.5B$_{pos}$.
The field strength in the dense core region of L1157 has been reported to be 1.3 - 3.5 mG by \cite{2013ApJ...769L..15S} using their 1.3 mm dust continuum polarization observations. These values are order of two magnitudes higher than our measurements which suggest that the core has stronger magnetic fields than in the envelope of L1157.

\subsection{Correlations between magnetic field, filament and outflow directions in L1157}

Star formation process begins with the accumulation of matter from inter-cloud medium. In models where magnetic fields are dynamically important compared to the turbulence \citep[e.g., ][]{1987ARA&A..25...23S, 1993ApJ...417..220G, 1993ApJ...417..243G, 1998ApJ...502L.163T, 2003ApJ...599..351A, 2003ApJ...599..363A}, the accumulation of matter is controlled by the ICMF. The gas slides along the field lines \citep{1999ApJ...527..285B, 2014ApJ...789...37V} forming filamentary structures aligned perpendicular to the ICMF. It is in these filaments that cores are found to be forming \citep{2013ApJ...777L..33P, 2015A&A...584A..91K}. For the reason that the assembly of the matter is guided by the magnetic fields, the ICMF direction is expected to get preserved deep inside the cores \citep{2009ApJ...704..891L,  2014ApJS..213...13H, 2015Natur.520..518L} predicting the ICMF to become parallel to the core magnetic field (CMF). The local CMF within individual cores (subcritical) provides support against gravity preventing them from collapse and thus account for the low efficiency  of the star formation process \citep[e.g., ][]{2001ASPC..248..515M}. The neutral particles, coupled weakly to the ions and hence to the magnetic fields, can drift towards the center of the core enabling it to amass more material. The increasing central mass gradually increases the mass-to-magnetic flux ratio leading the core to become supercritical and driving it to collapse under gravity. Under the influence of gravity, the initial uniform magnetic field is expected to get dragged toward the center of the core forming an hourglass-shaped morphology \citep{1993ApJ...417..220G, 1993ApJ...417..243G}. As the collapse progresses, a pseudo-disk is expected to form at the central region with the symmetry axis of the pinching to become perpendicular to it. The protostellar object embedded deep inside the cores, continues to build up mass through accretion and simultaneously develops a bipolar outflow. As the initial cloud angular momentum is expected to get hierarchically transferred to the cores and eventually to the protostar, the rotation axis (of the core/accretion disk) is expected to become parallel to the ICMF and CMF \citep{2006ApJ...645.1227M, 2006ApJ...637L.105M} and perpendicular to the core minor axis and the filament structure. 

In the above framework of magnetic field mediated star formation, a number of observational signatures that can manifest the role played by the magnetic fields are recommended \citep[e.g., ][]{2009ApJ...704..891L,  2014ApJS..213...13H, 2015Natur.520..518L}. Some of these are (i) the relative orientations between (a) the ISMF and the CMF, (b) the ISMF and the long axis of the filament, (c) the ISMF, CMF and bipolar outflows, (d) the filament and the outflows, and (ii) an hour-glass morphology of the magnetic field at the core scale with the symmetry axis perpendicular to the major axis of flattened pseudodisk. We examined these relationships in L1157 which is not only successful in forming a star but is also at its earliest stages of star formation therefore the initial conditions that led this cloud to form star may still be conserved.

The polarization measurements of the region surrounding L1157-mm were carried out at 1.3 mm (resolution 1.2$\arcsec$-4.5$\arcsec$) and 350 $\mu$m (10$\arcsec$) \citep{2013ApJ...770..151C, 2013ApJ...769L..15S} with the aim to trace the magnetic field orientation at the inner regions of the core \citep{1984ApJ...284L..51H, 1995ASPC...73...45G, 1999A&A...344..668G}. The SHARP and the CARMA vectors with their standard deviation have been quoted as P = 0.7$\%$ $\pm$0.2$\%$, $\theta$ =142$\degr\pm9\degr$ and P = 3.8$\%$ $\pm$0.1$\%$, $\theta$ =147.8$\degr\pm0.8\degr$, respectively. We adopted an orientation of 145$\degr\pm9\degr$ for the CMF which is the mean value of the magnetic field directions inferred from the SHARP and the CARMA results. The magnetic field inferred from the 1.3 mm polarization measurements is shown in the inset in Fig. \ref{fig:dss_her_pol_C18O} (b) by overplotting the vectors on the WISE 12 $\mu$m image. The CARMA and SHARP polarization measurements were carried out at the scales of $\sim400-1500$ au and $\sim3500$ au (using the distance of 340 pc), respectively thus representing the CMF. The offset between the relative orientations of the ICMF inferred from the optical (and the \textit{Planck}) polarimetry and the CMF obtained from the SHARP and CARMA is  $\sim18\degr\pm14\degr$. The nearly parallel orientations of ICMF and CMF suggest that the CMF is anchored to the ICMF in L1157.  The near parallel magnetic field geometry from large to core scales indicates that the fields have not been disturbed by turbulent motions resulting from the collapse of material during the star formation process in L1157.

%******************************************
\begin{figure*}
\centering
  \includegraphics[width=\textwidth, height=10cm]{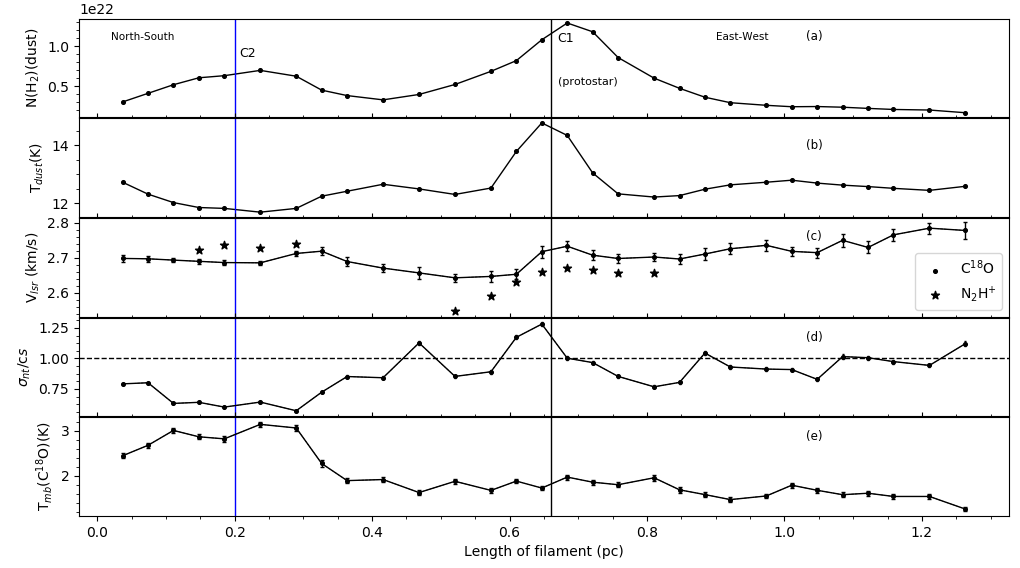}
\caption{Results of C$^{18}$O (1-0) and dust emission analysis along the cloud filament length. \textbf{(a)} Hydrogen column density derived using Herschel PACS and SPIRE images. \textbf{(b)} Dust temperatures. \textbf{(c)} Centroid velocity of \coo obtained from Gaussian fitting of profiles (filled dots). The centroid velocities obtained using hyperfine fitting of N$_{2}$H$^{+}$ (1-0) line coinciding with the positions along the filament are marked by filled stars. \textbf{(d)} Mach number which is the ratio of non-thermal velocity dispersion ($\sigma_{nt}$) along the line-of-sight and isothermal sound speed (c$_{s}$) at 10 K ($\sim$ 0.19 \kms). Blue line around 0.2 pc and black line 0.65 pc shows the position of clump C2 and class 0 protostar L1157-mm respectively. \textbf{(e)} Main beam brightness temperature using C$^{18}$O lines.}\label{fig:splot-c18o}
\end{figure*}
%******************************************

The orientation of the magnetic field towards L1157-IRS was found to exhibit an hour glass morphology \citep{2013ApJ...769L..15S}. The symmetry axis of the hourglass morphology is found to be perpendicular to a flattened structure ($\sim0.1-0.2$ pc) seen in absorption against a bright emission possibly due to the polycyclic aromatic hydrocarbons from the diffuse interstellar medium in the background \citep{2007ApJ...670L.131L}. The N$_{2}$H$^{+}$ emission, having a spatial extent of $\sim30,000$ au and oriented 75\dgr~with respect to the north to east \citep[elongated perpendicular to the outflow; ][]{2010ApJ...709..470C}, is found to coincide with the absorption feature. These structures extend further, at least, to the western side of L1157-mm coinciding with the east-west segment of the filament to upto a length of $\sim0.5$ pc. From the \textit{Filfinder} algorithm, the east-west segment of the filament is oriented at an angle of 79\dgr~(consistent with the orientation of N$_{2}$H$^{+}$ emission) is making an offset of 48\dgr~ with respect to the ICMF and 66\dgr~with respect to the CMF direction. The offset of 66\dgr~of CMF with respect to the east-west segment of the filament suggests that the star formation in L1157 supports a scenario where the magnetic field is sufficiently strong enough to have influenced the formation of at least the east-west segment of the filament structure  \citep{2013ApJ...774..128S, 2016A&A...586A.138P}. An alternate possibility for the formation of the east-west segment is discussed in section \ref{sec:motion}. The north-south segment of the filament, however, is found to be almost parallel to both the ICMF and CMF. 

The relative orientation of the bipolar outflow and the filament (the absorption feature and the N$_{2}$H$^{+}$ emission) in L1157 is found to be 82$\degr\pm10\degr$ \citep[assuming an uncertainty of $10\degr$ in the determination of the outflow position angles, e.g., ][]{2015A&A...573A..34S} which is almost perpendicular to each other. If we consider the outflow direction as a proxy to the rotation axis, then the orientation of outflow with respect to the filament provides evidence for the manner in which matter got accumulated prior to the initiation of the star formation. In L1157, the material might have got accumulated first onto the filament channeled by the  magnetic field lines aligned perpendicular to it and then as the density increased, the flow pattern changed its direction and might have flown along the filament towards the core. In such a scenario, one would expect the rotation of the protostar to be perpendicular to the filament since the local spin motion depends on the flow direction \citep{2017MNRAS.468.2489C, 2017ApJ...846...16S}. 

The offsets between ICMF and the outflow direction and between CMF and the outflow direction is $34\degr\pm16\degr$~ and $16\degr\pm14\degr$~\citep[see also] []{2013ApJ...769L..15S}, respectively which suggest that CMF is relatively more aligned with the outflow than the ICMF. Examples of both, alignment and misalignment or random alignment of outflows and magnetic field exist in the literature. Though the studies have shown that the outflows are preferentially misaligned or are randomly aligned with the magnetic fields, L1157-mm shows a good alignment especially between the outflow and CMF directions \citep[see also ][]{2013ApJ...768..159H}. Misalignment between magnetic fields and outflows are suggested as an essential condition to allow the formation of circumstellar disk \citep{2013ApJ...767L..11K}.
 
\subsection{Properties of the matter along the filament}\label{sec:fil_gas}

%**********************************************
\begin{figure*}
\centering
   \includegraphics[width=20cm, height=11cm]{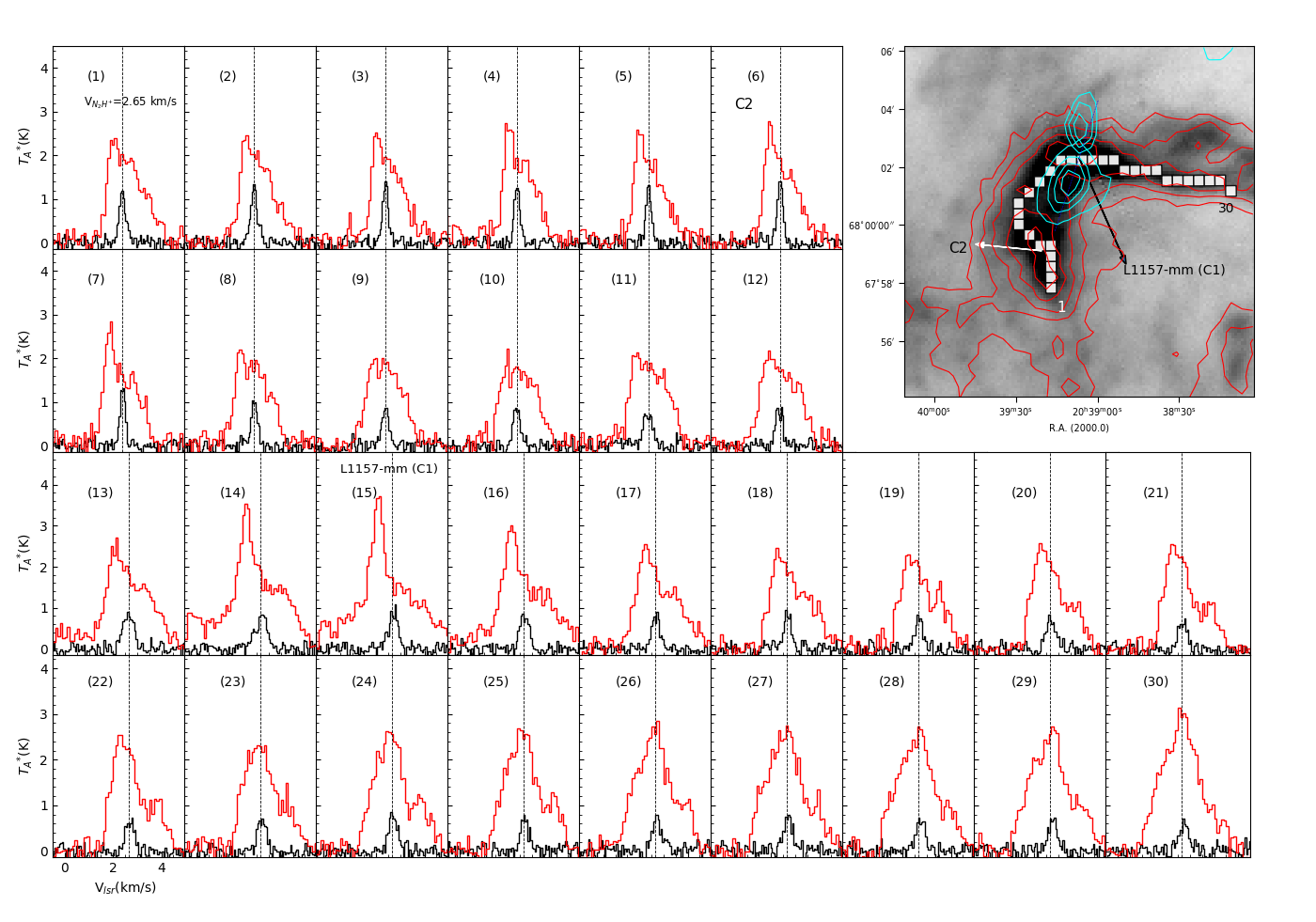}
  \caption{Distribution of \coo (black) and $^{12}$CO  (red) (1-0) line along the positions of \textit{Filfinder} skeleton. The dotted line shows the v$_{lsr}$ of the cloud which is adopted from \nthp peak velocity. The spatial positions of profiles are shown with filled white boxes on 250 $\mu$m \textit{Herschel} image (upper-right) from 1-30 starting from extreme south towards west along the filament. The protostar L1157-mm and clump C2 are marked by black and white arrows, respectively. Integrated intensity contours of C$^{18}$O (1-0) shown in red color are obtained by summing flux over the velocity intervals from 2.2-3.0 \kms. The contours start from 6$\sigma$ with intervals of 4$\sigma$ where $\sigma\sim$ 0.019 K \kms. The contours in cyan show the blue-shifted (towards north) and red-shifted lobe (towards south) of bipolar outflow. The levels for blue-shifted lobe range from 0.12-0.5 in steps of 0.08 K \kms and red-shifted lobe in range of 3.9-6.9 K \kms in intervals of 1 K \kms. The \co line was integrated from -2.2 to +2.3 \kms for high velocity wings in southern lobe and 3.0 to 3.9 \kms in northern lobe.}\label{fig:profiles-fil}
\end{figure*}
%**********************************************
To study the dust properties and large scale velocity field of the gas lying along the spine of the filament identified towards L1157, we derived peak line temperature, velocity centroids and velocity dispersion by making Gaussian fits to all \coo spectra having a S/N$\geqslant$3. The results are presented in Fig. \ref{fig:splot-c18o} as a function of the positions along the main axis of the filament. The southern end of the north-south segment of the filament is taken as the starting point. The positions of the two condensations C1, C2 identified using the \textit{Clumpfind} routine are marked. To compare the properties of dust and gas emission in the cloud, we convolved and regridded the  dust column density map at $\sim$36$\arcsec$ grid using the beam size of TRAO ($\sim49\arcsec$ at 110.20 GHz) (Jeong et al. 2019 submitted). The advantage of this approach is that the comparison is made over the same area on the source, but the disadvantage is that all the spatial structures smaller than this beam size get smoothed out. 

In the panels (a) and (b) of Fig. \ref{fig:splot-c18o}, we show the derived dust column density and the dust temperature calculated using the \textit{Herschel} data along the filament. The column density ranges from $1.7\times$ 10$^{21}$ to $1.3\times$ 10$^{22}$ cm$^{-2}$. The average dust column density of the filament is $\sim5\times$ 10$^{21}$ cm$^{-2}$. The dust temperature range from 12 to 15 K with an average temperature of 13 K. The column density and the dust temperature both peak at the position of L1157-mm (identified using thick line in black). However, the values remain nearly constant on both north-south and east-west segments of the filament. The peak value of the dust column density on the filament, 1.3$\times$ 10$^{22}$ cm$^{-2}$, is found at the position of C1 where L1157-mm is embedded. The mass-per-unit length values along the length of the filament were derived using the radial profile analysis as shown in Fig. \ref{fig:radfil-sub} (right lower panel). \cite{1964ApJ...140.1056O} considered the mass-per-unit length or line mass of an isothermal filament in equilibrium as 2c$_{s}^{2}$/G $\sim$ 15 M$_{\odot}$ pc$^{-1}$ at 10 K. The mass-per-unit length along the filament in L1157 ranges from 4-38 M$_{\odot}$ pc$^{-1}$. The line mass around C1 and C2 are larger than the equilibrium value which implies that these parts of the filament are supercritical. The values obtained around C1 and C2 are consistent with those found towards the Taurus molecular cloud \citep{2013A&A...550A..38P}. 

The mean column density profile of L1157 cloud is described with a Plummer-like function with a power law index of p$\sim$3 (See \ref{sec:fil}). The value falls in a range of the typical values of p ($\sim1.5-3$) obtained in the case of several filaments such as p$\sim2.7-3.4$ in L1517 \citep{2011A&A...533A..34H}, and p = 3 in L1495 \citep{2015A&A...574A.104T}. The radial equilibrium of the filamentary clouds can be explained by considering them as isothermal cylinders using pure hydrostatic models \citep{1964ApJ...140.1056O} or magnetohydrostatic models \citep{2000MNRAS.311...85F}. The former models can lead to density profiles as $\rho$ $\sim$ r$^{-4}$ and the latter ones to $\rho$ $\sim$ r$^{-2}$. Our results with power-law index, p$\sim$3 suggest that the filament column density profile supports a theoretical model with magnetic fields. The results of magnetic field studies presented here support the fact that the magnetic field has played an important role in the dynamical evolution of L1157.

 The \coo is detected at points all along the $\sim$1.22 pc length of the filament and correlates well with the dust emission. The distribution of \co (in red) and \coo (in black) profiles all  along the spine of the filament is shown in Fig. \ref{fig:profiles-fil}. Position 1 corresponds to the southern end of the north-west segment and position 30 corresponds to the western end of the east-west segment of the filament. There is a strong tendency that blue asymmetry in \co line profile is seen at high column density region while no clear asymmetry in those lines is seen at low column density region. The \co line profiles at positions 1-3 in the north-south segment of the filament show a blue asymmetry. As we approach towards the C2 (positions 4-7), the \co profile becomes blue-red asymmetry with the blue peak brighter than the red peak. Then the line profile becomes more of a symmetric one till the position 12. The profile becomes blue-red asymmetry again as we approach C1 and continues till the position 17. The \coo, which is an optically thin tracer, peaks at the velocity of the self-absorption, suggesting that the double-peaked profiles of \co (Fig. \ref{fig:profiles-fil} in the panels 4-7 and 13-17 are most likely due to inward motions assuming that the gas in the inner parts of the C1 and C2 has a warmer excitation temperature than that towards the envelope. The inward motions seen here could be interpreted as due to collapse or infall motion \citep{1999ApJ...526..788L,2001ApJS..136..703L,1998ApJ...504..900T}. The linear extent of the infall motion is around $\sim0.15$ pc. The \coo lines are well fitted with a Gaussian profile throughout the filament. The \coo line-width around the clump C2 is found to be narrower than that found around the core C1. The \co lines at positions 23-30 peaks at the systematic velocity of the cloud with an additional red component which is most likely caused due to the effect of the red lobe of the outflow.
 
 The values of the observed line-width (full-width at half maximum) of \coo profiles range from $\sim0.3$ to $0.6$ \kms. We fitted a single Gaussian profile to all the \coo profiles and obtained the peak velocites. The mean value of V$_{lsr}$ of the full L1157 filament was found to be 2.65$\pm$0.05 \kms. The systematic velocity of the cloud was estimated by taking an average of the velocities at positions where significant emission in the \nthp line was detected. The variation in the centroid velocity (V$_{lsr}$) all along the spine of the filament is shown in Fig. \ref{fig:splot-c18o} (c). The peak velocity changes from $2.64$ to $2.78$ km~s$^{-1}$ (2-3 velocity channels). The filament appears to be velocity coherent as there is no significant change in the peak velocity. This is consistent with the previous studies of nearby filaments forming low mass stars \citep{2011A&A...533A..34H}. The mean dispersion in the centroid velocities obtained from the Gaussian fit is  found to be $\sim0.03$ \kms. To the west of the C1, the east-west segment of the filament shows almost a constant value of V$_{lsr}$ ($\approx2.7$ \kms). Compared to this, the velocity structure of the north-south branch shows discernible variation. No notable variation in the values of V$_{lsr}$ obtained from \coo lines is seen along the filament except at the position where the north-west segment of the filament changes its direction towards the east-west segment. 
 
 The \nthp (1-0) line emission was detected towards both C1 and C2 with emission being prominent in regions around C1. The seven components in \nthp (1-0) spectra were simultaneously fitted with seven Gaussian forms at once with their line parameters given by \cite{1995ApJ...455L..77C}. We obtained V$_{lsr}$ of the cloud, line-width, total optical depth of all the components using fitting results. The peak velocity in \nthp lines varies from 2.54-2.74 \kms. The peak velocities obtained from the \nthp line towards C1 and C2 along the filament spine are shown in Fig. \ref{fig:splot-c18o} (c) using star symbols. The V$_{lsr}$ of the C1 and C2 was estimated as 2.62 \kms and for 2.72 \kms respectively. The \nthp line shows a systematic change in the velocity across the position of L1157-mm. The velocity gradient estimated using \nthp emission is found to be 0.37 \kms pc$^{-1}$. \citet{2012ApJ...756..168C} detected \nthp emission across an elongated region of $\sim30,000$ au (considering a distance of 340 pc) which is consistent with the flattened structure seen by \citet{2007ApJ...670L.131L}. Systematic variations in the velocity are seen in \nthp emission across the flattened envelope at $\sim30,000$ au \citep{2012ApJ...756..168C}, similar to the variations noticed by us at $\sim0.1$ pc scale. This suggests that the variations in the velocity observed at different scales \citep{2012ApJ...756..168C, 2015ApJ...814...43K} are most likely inherited from the bulk motion of the gas at the cloud scale.   
 
 The linewidth of a spectral line is a combination of thermal and non-thermal motions \citep{1983ApJ...270..105M}. Non thermal motions are generally arising from  turbulence in cloud or core scale mechanisms. We have separated out the thermal component from the observed line-width obtained from the Gaussian fitting analysis with the assumption that the two components are independent of each other. The non-thermal component is calculated as
 \begin{equation} 
\sigma_{nt} = \sqrt{(\sigma^{obs})^{2} - (\sigma^{th})^{2}}
\end{equation}
where $\sigma^{th}$ is $\sqrt{kT/\mu m_{obs}}$, thermal velocity dispersion, $\mu$ is the molecular weight of the observed \coo molecule, T is the gas temperature and k is the Boltzmann constant. The mach number (\textit{M}) is defined as the ratio of non-thermal component ($\sigma_{nt}$) and isothermal sound speed (c$_{s}$) shown in the panel (d) of Fig. \ref{fig:splot-c18o}. The variation shows the extent of non-thermal motions distributed along the positions of the filament. We find that much of the gas in the filament is subsonic as \textit{M} $\leq$ 1. Only the region around the protostar shows the signature of turbulent motions as the $\sigma_{nt}\geq$ c$_{s}$. The \coo peak line temperature is plotted as a function of position in the panel (e) of Fig. \ref{fig:splot-c18o}. The intensity peak observed in \coo towards the C1 is found to be shifted from the peak intensity in dust emission. We believe that this could be because of the depletion of the \coo molecules from the gas phase at the high density regions. On the other hand, the \coo line intensity peaks in the vicinity of C2 (T$_{mb}\approx3.0$ K) and remains roughly constant at T$_{mb}$ of $1.0 - 3.0$ K all along the parts of the filament. The \coo emission is lesser at the position of C1 where dust emission is the brightest and \coo is comparable to other positions along the east-west branch. Contrary to this, \coo emission is the strongest at the position of C2 where dust emission is second brightest. It might happen that \coo is highly depleted or photodissociated at region around C1. However, C2 clump seems less depleted in comparison with C1 core. This suggests that the C2 is chemically younger core than C1.
Overall the L1157 is velocity-coherent and mostly subsonic throughout its length though having internal dynamics around core C1 and C2. 

We derived total mass of the cloud by summing up all the N(H$_{2}$) values falling within the half-maximum contour level in the intensity map of \coo as shown in Fig. \ref{fig:profiles-fil} where it covers the high density region of the cloud. The corresponding pixels in the dust column density map were used to calculate the mass of cloud by dust emission. The mass of the L1157 cloud by gas emission was estimated as $\sim$ 8 M$_{\odot}$ whereas the mass of the cloud by dust emission was calculated as 16 M$_{\odot}$. There is a difference by a factor of more than 2 between mass (M(H$_{2}$)) calculated from \coo observations and \textit{Herschel} dust emission map. We expect the coupling of gas and dust in the interstellar matter at the volume densities of $\sim$ 10$^{5}$ cm$^{-3}$ which do not correspond to the critical density of \coo molecules \citep{2001ApJ...557..736G}. The typical uncertainty in the estimation of the M(H$_{2}$) value using dust emission is a factor of 2. The dominant factor contributing to the error in the mass estimation is the uncertainty in the opacity law. This value is an upper limit as we lack information on the inclination of the filament. The difference between the M(H$_{2}$) values derived from gas and dust can be attributed to various factors. The \coo molecules can deplete and freeze out on dust grains in high density regions ($n_{H_2} \geq 10^{5}$ cm$^{-3}$) and low temperatures (T $\leq$ 20 K). In addition, \textit{Herschel} is capable of tracing the dust column where the temperatures are higher but \coo line might get affected by the interstellar radiation field (ISRF) and get photodissociated at less dense regions of cloud (e.g., \citep{1999ApJ...523L.165C,2004A&A...416..191T,2016A&A...592L..11S}). The gas column density can change due to variation in CO-to-H$_{2}$ conversion factor or abundance ratio of optically thick \co and optically thin \coo tracers according to metallicity and column density gradients \citep{2010ApJ...721..686P,2013ARA&A..51..207B}.

\subsection{Physical parameters of the clump C2}

\begin{figure}
\centering
  \includegraphics[width=9.6cm, height=8.6cm]{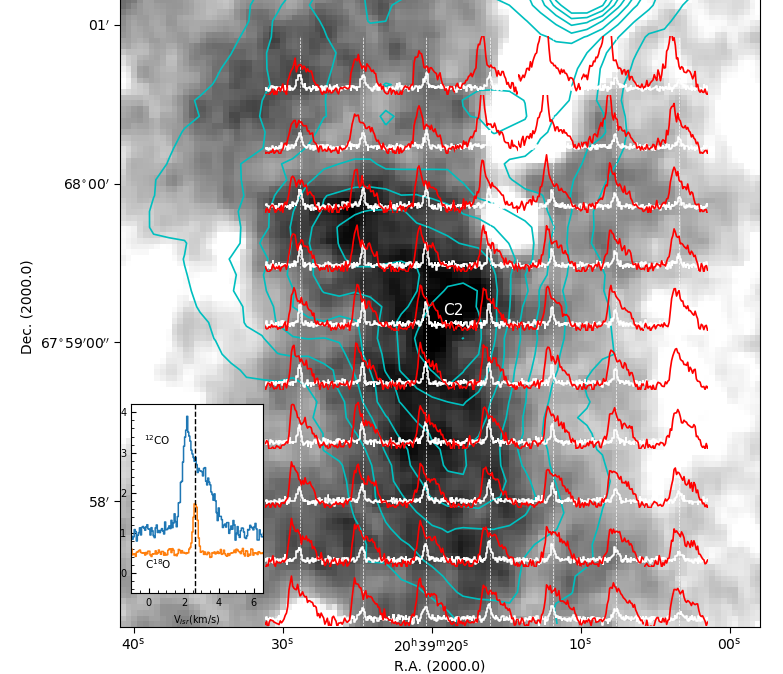}
  \caption{\co (red) and \coo (white) profiles overlayed on 12 $\mu$m \textit{WISE} emission map with C2 marked. The contours in cyan color show the 250 $\mu$m dust intensity emission and the contour levels are in range 50-120 MJy/sr in steps of 10 MJy/sr. In the inset shown two profiles for \co (in blue) and \coo lines (in orange) averaged over half-maximum contour in intensity map of \nthp line. }\label{fig:profiles-c2}
\end{figure}

The C1, which is currently forming the protostar L1157-mm, shows supersonic turbulent motions in \coo lines. The dust temperature was found to be $\sim15$ K. Previous studies have already characterised the C1 using dust continum and line mapping observations at different spatial and spectral resolutions \citep{1997A&A...323..943G}. In this section we determined the properties of C2 to characterize its evolutionary state. The clump C2 has a peak dust temperature (T$_{d}$) $\sim$ 12 K and peak column density $\sim$ 9 $\times$ 10$^{21}$ cm$^{-2}$. As discussed earlier, there exist blue-red asymmetry in \co lines towards center of C2 and the \coo peaks at the velocity of the self-absorption indicating the presence of an inward motion (Fig. \ref{fig:profiles-fil} and Fig. \ref{fig:profiles-c2}). The lack of high-velocity wings in the line profile suggests that the region is not affected by the outflow though the southern edge of the blue-shifted lobe of the outflow spatially coincides with the outer periphery of C2. The outflow is almost in the plane-of-sky with the inclination angle of 10$\degr$ \citep{1996A&A...307..891G}. We quantified the outflow energetics with its mass and kinetic energy as 0.05 M$_{\odot}$ and 1.2 $\times$ 10$^{43}$ ergs, respectively. The sources of uncertainty in the estimation of the outflow parameters depend on the velocity boundary between the high velocity wings and the ambient velocity and the inclination angle with respect to the line-of-sight. The gravitational binding energy of the C2 is calculated as 0.53 $\times$ 10$^{43}$ ergs. Thus, the outflow has the potential to disturb the ambient gas and may affect C2 in future. The motion of the gas molecules around C2 is very quiescent with subsonic turbulence and high \coo intensity. Fig \ref{fig:splot-c18o} (c) shows that the subsonic non-thermal motions are mostly associated with the region around C2. The properties of prestellar or starless cores in low mass star forming regions like Taurus and Ophiuchus have been studied in detail \citep{2007prpl.conf...17D,Gregersen_2000,2002ApJ...575..950O,1998A&A...336..150M} and the values found here for C2 are consistent with the typical properties of starless core. 

The measure of non-thermal line-widths using molecular line diagnostics can be used to investigate whether the core is virially bound or not. We derived the virial mass of C2 using the averaged total velocity dispersion of \coo line. If the mass of the clump is less than the virial mass, the cloud is not gravitationally bound and may expand. We derived the virial mass using the formula \citep{1988ApJ...333..821M},
\begin{equation}
M_{vir} = \frac{k\sigma^{2}R}{G}  
\end{equation} where k depends on the density distribution, G is the Gravitational constant and R is the cloud radius. The total velocity dispersion is given by the equation,
%, $\rho$ = r$^{\alpha}$
\begin{equation}
\sigma = \sqrt{\frac{kT_{kin}}{\mu_{m_{H}}}+ \sigma^{2}_{nt}}
\end{equation}
Assuming a Gaussian velocity distribution and density profile distribution as $\rho$ = r$^{-2}$, we can express the above equation also in terms of solar mass as M$_{virial}$ = 126$\Delta$v$^{2}$R, where $\Delta$v is the FWHM velocity of \coo lines in km s$^{-1}$ along the line-of-sight and R, the radius in parsec units. By approximating C2 as an ellipse on the sky projection, the radius of the core was estimated as $\sqrt{FWHMx}\times\sqrt{FWHMy}$, where FWHMx and FWHMy are the full-width at half maximum diameters along major and minor axis, respectively. The value of average FWHM velocity of the gas was calculated as 0.5 \kms. We calculated the effective radius of C2 as $\sim$ 0.1 pc using FWHM of major axis $\sim$ 1.3 $^{'}$ and minor axis $\sim$ 0.7 $^{'}$. The virial mass of C2 was estimated as $\approx$ 3.1 M$_{\odot}$. The major contribution of uncertainty in the calculation of the virial mass comes from the uncertainty in the distance estimation which is $\sim1$\%. There is a variation of 5\% in the value of virial mass due to distance uncertainty. We summed up all the pixels with column density values falling within the derived radius of clump C2 in moment zero map of C$^{18}$O line. The total gas mass of C2 is calculated as $\sim$ 2.5$\pm$0.3 M$_{\odot}$. We found that the clump C2 is on the verge of being gravitationally bound since M$_{virial}$ $\approx$ M$_{gas}$. The dust mass around clump C2 is calculated as $\approx$ 5 M$_{\odot}$ using same region as used in the calculation of gas mass.

%********************************************************************
%********************************************************************

\begin{figure}
\includegraphics[width=8.1cm, height=7.0cm]{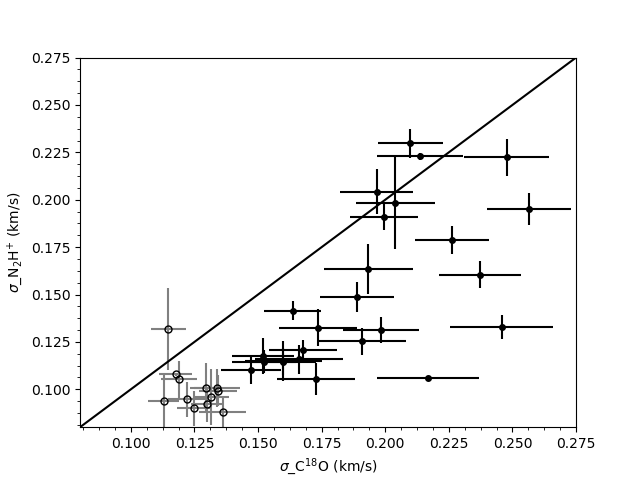}
\caption{Variation of velocity dispersion measured in \coo and \nthp profiles. The filled circles are the points near C1 and unfilled circles are around C2.}\label{fig:sig-n2hp-c18o}
\end{figure}

The \nthp line traces the dynamics of the dense central part of the core while the \coo line traces the dynamics of the surrounding less dense material in the envelope. We studied the core-to-envelope motion around C1 and C2 by comparing the centroid velocity of \nthp and \coo lines and the velocity dispersion. 
As can be noticed, the velocities of the different tracers match with each other as shown in Fig \ref{fig:splot-c18o} also. The average difference in centroid velocities of \coo and \nthp around clump C2 is 0.03$\pm$0.02 \kms and around core C1 is 0.06$\pm$0.02 \kms which means that the velocities of the different tracers differ on average by less than one fifth or one third of the sound speed. This good match rules out any significant relative motions between different density regimes of
the gas and, in particular, rules out any possibility of systematic drift between the dense cores (traced by \nthp) and the surrounding gas (traced by \coo). This result is in good agreement with the previous studies that probed relative motion between the dense inner region and envelope of the cores \citep{2007ApJ...668.1042K,2004ApJ...614..194W,2007MNRAS.374.1198A}. The line width obtained in \coo line is found to be relatively broader (by 1-2 channels around core C1) when compared with that of \nthp (1-0) as shown in Fig \ref{fig:sig-n2hp-c18o}. The difference in line-width around core C1 is within a channel spacing for both the tracers. The motions in C2 is subsonic in both \coo and \nthp lines as compared to that of C1. This behaviour of line broadening is consistent with the previous studies where starless cores are expected to have less turbulent motions and protostellar cores show more broader line-width \citep{2007ApJ...668.1042K}.

The \nthp and \coo lines can be used to find out the extent of chemical evolution of the dense cores. \nthp can only form in significant amounts after \coo freezes out on dust grains since both the molecules form by competing reactions \citep{1999ApJ...523L.165C}. \nthp is observed to be good tracer of gas at densities $\sim$ 10$^{5}$-10$^{6}$ cm$^{-3}$ whereas \coo will deplete at these densities \citep{2002ApJ...569..815T}. At later stages of evolution when the central protostar formation has taken place, the \nthp will get destroyed due to rise in temperature and CO molecules will start forming. We used the integrated intensity of \coo and \nthp lines to calculate the ratio of intensities. If this ratio, \textit{R} \textgreater 1, it implies that the core has not evolved to the extent that the carbon molecules could freeze out on the surface of dust grains and therefore, chemically young. We averaged the intensity values around the starless core C2 where significant ($\geq$ 3$\sigma$) emission of \nthp is obtained. The ratio is found to be greater than 1 which implies that the core has not yet chemically evolved. C1 core shows highly enhanced distribution of \nthp line. This may be due to the significant depletion or photodissociation of \coo molecules in the core which usually play a role for the destruction of \nthp \citep{1999ApJ...523L.165C}. \nthp in C2 core is also enhanced, but not as much as C1 core. This can be also explained by overabundance of \coo molecules than C1 core.
%**********************************************
%**********************************************

\subsection{Motion of L1147/1158 and L1172/1174 complexes}\label{sec:motion}
%********************************************************************
\begin{figure}
\centering
  \includegraphics[width=8.5cm, height=8cm]{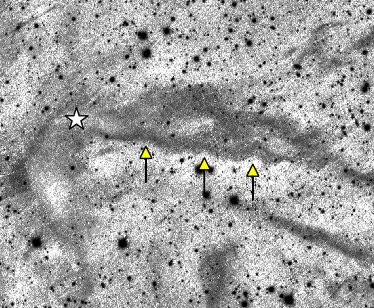}
  \caption{The PanSTARRs-z band image of the field containing L1157. The sinuous features are identified with arrows and the position of the protostar L1157-IRS is identified using a star symbol. The contrast of the original image is adjusted to reveal the features.}\label{fig:sinuous}
\end{figure}
%********************************************************************

A PanSTARRs z-band ($\lambda_{eff}=8668.5$\AA) image \citep{2016arXiv161205560C} of the field containing L1157 (contrast of the original image is adjusted to highlight the features) in the Galactic coordinate system is shown in Fig. \ref{fig:sinuous}. The shape of the southern boundary of the east-west segment is found to be sinuous by the arrows. \citet{1987ApJ...318..702O} and \citet{1988ApJ...325..320O} have catalogued a number of high-latitude cometary/filamentary objects gleaned from the IRAS images and suggested that their morphologies can be explained as a result of cloud-ISM interaction. A flow of velocity of V around an object of length L can be characterized using the Reynolds number (R$_{e}$=$\rho$LV/$\mu$), where $\rho$ is the ambient fluid density, and $\mu$ is the fluid viscosity. While flows of extremely low ($\approx$10) and intermediate ($\approx$50) values of R$_{e}$ are expected to show a smooth laminar flow pattern and an irregular structures and vortices respectively, higher values of R$_{e}$($\gg$100) is expected to produce a fully turbulent flow. \citet{1988ApJ...325..320O} suggested that for a relatively low value of R$_{e}$ ($\lesssim$10), the mass lost by ablation due to the motion of a cloud through ambient medium can form long sinuous filaments.

%\begin{comment}
%********************************************************************
\begin{figure}
\centering
  \includegraphics[width=8.4cm, height=8cm]{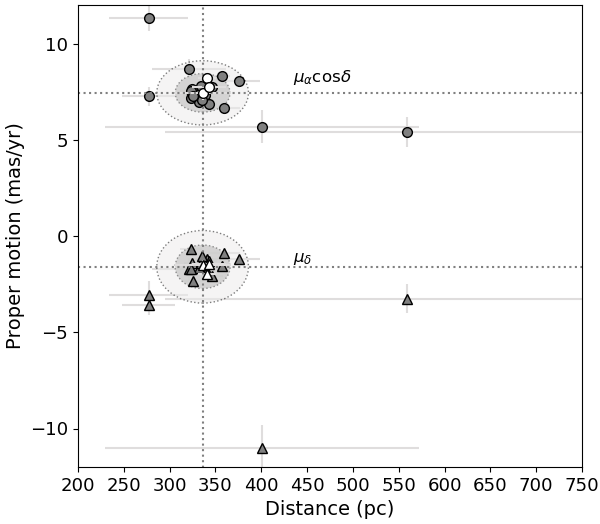}
  \caption{Proper motion values and distances of the 23 YSO candidates associated with L1147/1158 and L1172/1174 complexes. The darker and lighter shaded ellipses are drawn using 3 and 5 times the median absolute deviation values of the distance and the proper motions respectively.}\label{fig:pm_dist}
\end{figure}
%********************************************************************
%\end{comment}
To calculate the space velocity of L1147/1158 complex, required to estimate the value of R$_{e}$, we made use of the proper motion values of the YSOs associated with the region and the radial velocities of the clouds. Because L1147/1158 ($l\sim102.2$\dgr, $b\sim+15.3$\dgr) and L1172/1174 ($l\sim104.03$\dgr, $b\sim+14.3$\dgr) complexes are located close to each other ($\sim$2\dgr) in projection and share similar radial velocities  \citep{1983ApJ...264..517M, 1997ApJS..110...21Y}, we included the YSOs associated with L1172/1174 also in our analysis. A total of 58 YSO candidates are identified so far in the vicinity of L1172/1174 \citep{2009ApJS..185..451K, 2009ApJS..185..198K, 2013MNRAS.429..954Y}. We found a \textit{Gaia DR2} counterpart for 20 of them well within a search radius of 1$\arcsec$. As in the case of L1147/1158, we obtained their distances from the \citet{2018AJ....156...58B} catalog and $\mu_{\alpha\star}$ and $\mu_{\delta}$ values from the \citet{2018A&A...616A...1G} catalog. Again, we considered only those sources having their parallax and proper motion values greater than or equal to three times of the corresponding errors. The \textit{Gaia DR2} results are presented in Table \ref{tab:yso_dist_pm} and shown in Fig \ref{fig:pm_dist}. The 3 YSOs in L1147/1158 complex (discussed in section \ref{sec:dist}) are shown using open circles ($\mu_{\alpha\star}$) and triangles ($\mu_{\delta}$) while the 20 YSO candidates in the L1172/1174 complex are shown using closed circles ($\mu_{\alpha\star}$) and triangles ($\mu_{\delta}$). Of the 20 sources in L1172/1174 complex, 15 of them are clustered along with the 3 sources in L1147/1158 complex. The median (median absolute deviation) values of the distance, $\mu_{\alpha\star}$ and $\mu_{\delta}$ for sources in L1172/1174 complex are 335 (11) pc, 7.301 (0.386) mas/yr and -1.619 (0.427) mas/yr, respectively. Previous distance estimate to L1172/1174 was $288\pm25$ pc \citep{1992BaltA...1..149S}. When combined, the YSO candidates from both the complexes together, we obtained a median (median absolute deviation) values of the distance, $\mu_{\alpha\star}$ and $\mu_{\delta}$ as 336 (10) pc, 7.436 (0.334) mas/yr and -1.599 (0.377) mas/yr, respectively. 

The darker shaded ellipses in Fig \ref{fig:pm_dist} are drawn using 3 times the median absolute deviation values of distance and proper motion. All the 3 and the 12 YSOs associated with L1147/1158 and L1172/1174, respectively are found within the distance-$\mu_{\alpha\star}$ ellipse. Four more YSOs associated with L1172/1174 could be included if we consider the ellipses drawn with 5 times the median absolute deviation values of distance and $\mu_{\alpha\star}$. Similarly, all the 3 and the 14 YSOs associated with L1147/1158 and L1172/1174, respectively are found within the distance-$\mu_{\delta}$ ellipse. Two more from L1172/1174 get added if we consider the ellipses drawn with 5 times the median absolute deviation values of distance and $\mu_{\delta}$. Four sources are found to be clear outliers. The lighter shaded ellipses in Fig \ref{fig:pm_dist} are drawn using 5 times the median absolute deviation values of distance and proper motion. All the sources found within these ellipses are considered as part of L1147/1158 and L1172/1174 complexes and included in our analysis. The results imply that the two complexes are related to each other both spatially and kinematically. 

\begin{figure}
   \includegraphics[width=8.7cm, height=8.5cm]{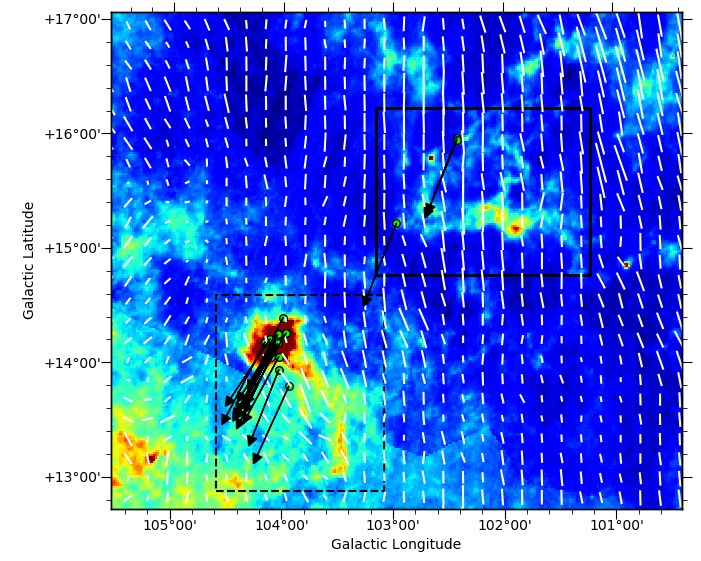}
  \caption{Planck polarization vectors in white overplotted over color scale \textit{AKARI} 160 $\mu$m emission map of L1147/L1158 complex. Proper motion values and distances of 19 YSO candidates associated with L1147/1158 and L1172/1174 complexes. The box in bold lines shows the L1147/1158 complex and in dashed lines shows L1172/1174 complex.}\label{fig:pol-plk-gal}
\end{figure}

The proper motions of the sources measured by the \textit{Gaia} are in the equatorial system of coordinates. To understand the motion of objects in the Galaxy, we need to transform the proper motion values from the equatorial to the Galactic coordinate system $\mu_{l\star}=\mu_{l}$ cos$b$ and $\mu_{b}$. We transformed the proper motion values using the expression \citep{2013arXiv1306.2945P},
\begin{equation}
\left[\begin{array}{c}\mu_{l\star}\\\mu_{b}\end{array}\right]=\frac{1}{cosb}\begin{bmatrix}\text{C}_{1} & \text{C}_{2}\\-\text{C}_{2} & \text{C}_{1}\end{bmatrix}\left[\begin{array}{c}\mu_{\alpha\star}\\\mu_{\delta}\end{array}\right]
\end{equation}
where the term $cosb$ = $\sqrt{\text{C}_{1}^{2} + \text{C}_{2}^{2}}$ and the coefficients C$_{1}$ and C$_{2}$ are given as,
\begin{equation}
\begin{array}{l@{}l}
\text{C}_{1} = sin\delta_{G}~cos\delta - cos\delta_{G}~sin\delta~cos(\alpha-\alpha_G) \\
\text{C}_{2} = cos\delta_{G}~sin(\alpha-\alpha_G)
\end{array}
\end{equation}
The equatorial coordinates ($\alpha_{G}$, $\delta_{G}$) of the North Galactic Pole are taken as 192\dgr.85948 and 27\dgr.12825, respectively \citep{2013arXiv1306.2945P}. The proper motion of the YSO candidates in the Galactic coordinates are drawn in Fig. \ref{fig:pol-plk-gal} using arrows in black. The mean values of $\mu_{l\star}$ and $\mu_{b}$ are found to be 3.499 mas/yr and $-6.815$ mas/yr, respectively, and the corresponding proper motion position angle is found to be 153\degr~. The arrows show the sense of the motion of the sources on the sky plane. If we assume that the cloud and the YSO candidates are expected to share similar kinematics as a result of them being born inside the cloud, then the arrows should also represent the motion of the clouds on the sky plane. Presence of reflection nebulosity around a number of these YSO candidates provide evidence of their clear association with the cloud. 

Based on our N$_{2}$H$^{+}$ observations, the V$_{lsr}$ velocity of L1157 is found to be $+2.65$ km~s$^{-1}$. \cite{1997ApJS..110...21Y} carried out molecular line survey in the direction of the Cepheus region which include L1148/L1157 and L1172/1174 cloud complexes. They found about nine clouds that showed the V$_{lsr}$ velocity in the range of $+2.6$ km~s$^{-1}$ to $+4.8$ km~s$^{-1}$. The mean value of the V$_{lsr}$ velocities of these nine clouds is found to be $\sim3.0$ km~s$^{-1}$. We took this value as the radial velocity of both the complexes. We converted this value from the LSR system to the heliocentric system as V$_{r}=-10.75$ km~s$^{-1}$. The two tangential velocity components along the Galactic longitude and latitude are calculated using V$_{l} = 4.74d\times\mu_{l\star}$ and V$_{b} = 4.74d\times\mu_{b}$, respectively. The factor 4.74 is the ratio of the au expressed in kilometer and the number of seconds in a tropical year. The $d$ is the distance in parsec of the individual stars obtained from the \citet{2018AJ....156...58B} catalog. Then we calculated the velocities U, V, W directed along the rectangular Galactic coordinate axes using the expressions \citep[e.g., ][]{2019AstL...45..109B},
\begin{equation}
\left[\begin{array}{c}U\\V\\W\end{array}\right]=\begin{bmatrix}cosl~cosb & -sinl &-cosl~sinb \\sinl~cosb & cosl & -sinl~sinb\\ sinb & 0 & cosb\end{bmatrix}\left[\begin{array}{c}\text{V}_{r}\\\text{V}_{l}\\\text{V}_{b}\end{array}\right]
\end{equation}
The velocity U is directed from the Sun toward the Galactic center with the positive direction being toward the Galactic center, V is positive in the direction of Galactic rotation, and W is positive directed to the north Galactic pole. The mean values of (U, V, W) for the 19 YSO candidates toward L1147/L1158 and L1172/1174 cloud complexes are ($-3.6$, $-8.8$, $-13.3$) km~s$^{-1}$ with a standard deviation of (0.6, 0.3, 0.9) km~s$^{-1}$. To determine the motion of the complexes with respect to the Galactic frame of reference, we transformed the heliocentric velocities to the LSR velocities by subtracting the motion of the Sun with respect to the LSR from the heliocentric velocities. In a number of studies \citep[e.g., ][]{2019MNRAS.484.3291T, 2019A&A...621A..48L} the velocity components of the motion of the Sun with respect to the LSR estimated by \citet{2010MNRAS.403.1829S} of (U, V, W)$_{\odot}$ = (11.1$^{+0.69}_{-0.75}$, 12.24$^{+0.47}_{-0.47}$, 7.25$^{+0.37}_{-0.36}$) km~s$^{-1}$ are used. However, recently, based on the \textit{Gaia} DR2 data, the velocity components of the Sun's motion have been re-evaluated using, for example, stars \citep{2019ApJ...872..205L, 2019RAA....19...68D}, open star clusters \citep{2019AstL...45..109B}, OB star samples \citep{2018AstL...44..676B} and white dwarfs \citep{2019MNRAS.484.3544R}. We used the most recent values of (U, V, W)$_{\odot}$ = (7.88, 11.17, 8.28)$\pm$(0.48, 0.63, 0.45) km~s$^{-1}$ obtained by \citet{2019AstL...45..109B} for the transformation from the heliocentric to the LSR velocities. The mean value of (u, v, w) estimated for the 19 YSO candidates toward L1148/L1157 and L1172/1174 cloud complexes is found to be (4.3, $-2.9$, $-5.0$) km~s$^{-1}$ with a standard deviation of (0.6, 0.3, 0.9) km~s$^{-1}$. The results imply that the L1148/L1157 and L1172/1174 cloud complexes are collectively moving in the direction of Galactic center, opposite to the Galactic rotation and coming towards the Galactic plane.  

\begin{figure*}
\centering
  \includegraphics[width=12cm, height=11cm]{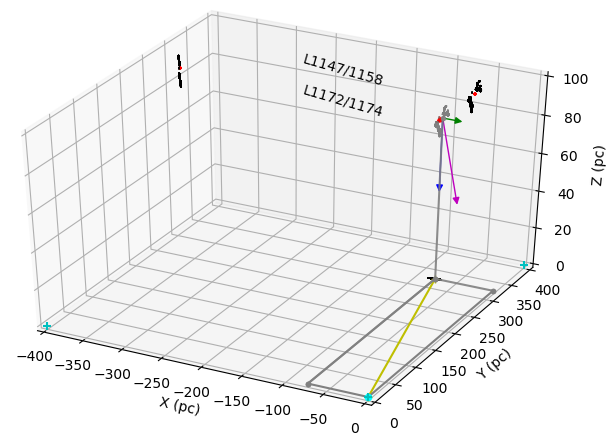}
  \caption{The 3-D motion of L1147/1158 and L1172/1174 complexes is presented on a rectangular coordinate system with the Sun as the origin. The X-axis, Y-axis, and Z-axis run parallel to the Sun-Galactic center vector, perpendicular to the Sun-Galactic center vector and perpendicular to the Galactic plane respectively. The positive direction of X-, Y- and Z-axis is the direction towards the Galactic center, in the direction of Galactic rotation and towards the Galactic North pole respectively. The arrows in green, red and blue represent the components of velocity in the X-, Y- and Z-axis respectively. The arrow in magenta represents the resultant velocity of the complex. The line in yellow shows the projection of the distance of the complex on the Galactic plane. The scale of the vectors is taken as eight times the magnitude of velocity.}\label{fig:3d_velocity}
\end{figure*}

The motion of the complex is presented on a rectangular ($x, y, z$) co-ordinate system with the Sun as the origin as shown in Fig. \ref{fig:3d_velocity}. The $Ox$ axis runs parallel to the Sun-Galactic center direction, $Oy$ in the Galactic plane but perpendicular to the $Ox$ and $Oz$ is perpendicular to the Galactic plane. The positive direction in $Ox$, $Oy$ and $Oz$ being the direction toward the Galactic center, in the direction of the Galactic rotation and towards the Galactic North pole, respectively. We computed the position of the complexes (X, Y, Z) as (-80, 319, 85) pc. The resultant velocity of the complexes as a whole is found to be $\sim7$ km~s$^{-1}$. 

 The Planck magnetic field vectors from a region containing both the complexes are also shown in Fig. \ref{fig:pol-plk-gal}. Though there are regions where projected magnetic fields show a rotation in the position angles (towards L1172/1174 complex), as a whole, the projected field orientation towards the region is found to be almost parallel to the galactic latitude. The projected magnetic field direction obtained from the median value of the galactic polarization position angles is $\sim5$\degr~ from the north. This implies that the projected motion of the complexes is making an angle of 32\degr~ with respect to the projected magnetic field orientation. Following the procedure from \citet{1987ApJ...318..702O} and \citet{1988ApJ...325..320O}, we made a rough estimate of the value of R$_{e}$ for L1157 and found it to be $\sim3$. Major contribution of uncertainty in the calculation of R$_{e}$ comes from the large uncertainty in the estimation of the density and the viscosity of the ambient medium. We ignored the effects of magnetic field in our calculation as the offset between the projected magnetic field and the direction of the cloud's motion is $\sim30$\degr. Studies with various values of offsets between the orientation of the magnetic field lines and the direction of the cloud motion have shown that only for large enough offsets magnetic fields play a significant role in the dynamical evolution of the cloud \citep{1994ApJ...433..757M, 1996ApJ...473..365J, 1999ApJ...517..242M}. The number density of the ambient medium is estimated as 0.09 cm$^{-3}$ adopting a density of 0.17 cm$^{-3}$ along the galactic plane and an exponential scale height of 125 pc \citep{1988ApJ...325..320O}. Assuming that the cloud is in pressure equilibrium with the ambient medium, we estimated the temperature of the ambient medium as $\sim10^{6}$ K using the average values of the number density and temperature of the cloud calculated from the dust emission. We used V = 7 km~s$^{-1}$ and L= 1 pc in the calculation. The low value of R$_{e}$ is consistent with the smooth morphology of the cloud structure as depicted in Fig. \ref{fig:sinuous}. 

%********************************************************************
%********************************************************************
%********************************************************************
%********************************************************************
Presence of several shell structures like the Cepheus flare shell (CFS) which is an old supernova remnant and the Loop III which is a giant radio continuum feature \citep{2009ApJS..185..198K} are taken as signature of multiple supernova explosions that might have occurred towards the region. The complexes L1147/1158 and L1172/1174 are located outside the CFS but at the periphery of the Loop III \citep{2009ApJS..185..198K}. It is possible that these supernova events in the past might have transferred the material towards the high latitude regions and that now the material is moving downward towards the galactic plane. \citet{1991A&A...249..493H}, in their study of the entire L1147/1158 complex, noted that the south-eastern boundary of the complex shows a remarkably sharp edge parallel to the Galactic plane. Based on the spectra obtained along the latitude, $l$ = 102\dgr.73, they detected a velocity gradient and attributed it to cloud rotation with the angular velocity vector pointing perpendicular to the Galactic plane. Alternatively, the sharp boundary and the velocity gradient seen could also be due to the bulk motion of the cloud material. The star formation happening in the clouds associated with the complexes might be as a result of their interaction with the ambient medium as they travel.

\section{Conclusions}\label{sec:summary}
We present results of a study conducted on a molecular cloud L1157 which is part of a cloud complex L1147/1158. Currently, formation of a Class 0 protostar, L1157-mm, having a spectacular bipolar outflow is taking place. The extreme youth of the protostar implies that the initial conditions that guided the cloud to form a star may still be preserved. We made R-band polarimetry of the cloud to trace magnetic field geometry of the cloud. We also made observations in $^{12}$CO, C$^{18}$O and N$_{2}$H$^{+}$ ($\rm J=1-0$) lines to investigate the kinematics of the material associated with the cloud. The main results obtained are summarized below.

\begin{enumerate}
    \item We estimated distance to the L1147/1158 complex using the YSOs associated with the cloud complex. The distances of the YSOs are estimated based on the parallax measurements and the proper motion values obtained from the \textit{Gaia} DR2 database. The estimated distance is $340\pm3$ pc.
    \item We obtained polarization measurements of 62 stars projected towards the direction of L1157 (within a region of 0.3\dgr$\times$0.3\dgr field). Based on the degree of polarization vs. distance plot for the stars observed by us and those from the \citet{2000AJ....119..923H} catalog, we present additional evidence of the presence of the cloud at $\sim340$ pc distance.
    \item Using the \textit{Filfinder} algorithm on the dust column density map of L1157 obtained from the \textit{Herschel} data, we traced a filament which is found to be $\sim1.2$ pc in length and oriented at a PA of 79\dgr~(east-west segment). Near to the protostar, the filament changes its orientation and becomes almost perpendicular (north-south segment). Using the \textit{Radfil} algorithm, the average filament width is estimated as $\sim0.09$ pc and the radial distribution of the material is fitted with a Plummer-like density profile of power-law index of $p = 3$. Using the \textit{Clumpfind} algorithm we identified two cores (C1 and C2) that are found to be located on the filament. In one of these cores is where L1157-mm is currently embedded. 
    \item The ICMF traced by our R-band polarization measurements of the stars background to the cloud is found to be well ordered at $\sim0.2-2$ pc scale. The geometry of the ICMF inferred from the \textit{Planck} 353 GHz data was found to be in good agreement with our R-band polarization results. The strength of the magnetic field calculated based on our data is found to be $\sim50$ $\mu$G. The ICMF is oriented at a PA of 127\dgr$\pm$12\dgr. 
    \item Based on the relative orientations between the ISMF, CMF, filament and outflow and the presence of an hour-glass morphology of the magnetic field at the core scale with its symmetry axis orthogonal to the major axis of the flattened pseudodisk suggest that magnetic field has played an important role in the evolution of L1157 to become a star forming core.
    \item We made $^{12}$CO, C$^{18}$O and N$_{2}$H$^{+}$ line observations of the entire region covering L1157 cloud. The \coo is detected at points all along the $\sim1.2$ pc long filament and found to correlate well with the dust emission. A blue-red asymmetry is observed in $^{12}$CO towards both C1 and C2 with C$^{18}$O peaking at the systematic velocity of the cloud signifying infall motion of the material. We found no significant change in V$_{lsr}$ velocity along the filament except at the location where the north-south segment changes its direction towards the east-west segment of the filament. The N$_{2}$H$^{+}$ ($\rm J=1-0$) line also shows a systematic change in the velocity across C1 suggestive of the presence of a bulk motion in the gas. 
    \item The east-west segment of the filament presents a sinuous structure. It is believed that the sinuous features seen in clouds are believed to occur due to cloud-ISM interaction. The dynamical state of such interactions depends on the Reynolds number which is found to be $\sim3$ in L1157. For such a low value ($\lesssim10$) of Reynolds number, the cloud motion through ambient medium can cause mass loss by ablation and can form long sinuous filaments.
\end{enumerate}

\section*{Acknowledgements}
This work has made use of data from the European Space Agency (ESA) mission {\it Gaia} (\url{https://www.cosmos.esa.int/gaia}), processed by the {\it Gaia} Data Processing and Analysis Consortium (DPAC, \url{https://www.cosmos.esa.int/web/gaia/dpac/consortium}). Funding for the DPAC has been provided by national institutions, in particular the institutions participating in the {\it Gaia} Multilateral Agreement. The Planck Legacy Archive (PLA) contains all public products originating from the Planck mission, and we take the opportunity to thank ESA/Planck and the Planck collaboration for the same. CWL is supported by Basic Science Research Program through the National Research Foundation of Korea (NRF) funded by the Ministry of Education, Science and Technology (NRF-2019R1A2C1010851). A.S. acknowledges financial support from the NSF through grant AST-1715876. This research has made use of the SIMBAD database, operated at CDS, Strasbourg, France. We also used data provided by the SkyView which is developed with generous support from the NASA AISR and ADP programs (P.I. Thomas A. McGlynn) under the auspices of the High Energy Astrophysics Science Archive Research Center (HEASARC) at the NASA/ GSFC Astrophysics Science Division.

%********************************************************************8
\begin{table*}
\begin{small}
\caption{\textit{Gaia} results of YSOs associated with the L1147/1158 and L1172/1174 complexes.}\label{tab:yso_dist_pm}
\renewcommand{\arraystretch}{1.3}
\begin{tabular}{lccccccccc} \hline
Source Name				&RA             &Dec           & $l$& $b$      & distance   &$\mu_{\alpha}$ &$\Delta\mu_{\alpha}$ & $\mu_{\delta}$ &$\Delta\mu_{\delta}$\\
                        & ($^{\circ}$)  & ($^{\circ}$) & ($^{\circ}$)  & ($^{\circ}$) &(pc)& (mas/yr) & (mas/yr)& (mas/yr) & (mas/yr)\\
     (1)&(2)&(3)&(4)&(5)&(6)&(7)&(8)&(9)&(10)\\\hline\hline
     \multicolumn{10}{c}{L1147/1158 complex}\\
2MASS J20361165+6757093	&309.048672&	67.952608&	102.4221&	15.9738&	336$^{20}_{-22}$&	7.436&	0.351&	-1.514&	0.284 \\
IRAS 20359+6745			&309.082855&	67.942131&	102.4205&	15.9573&	343$^{12}_{-13}$&	7.753&	0.189&	-1.469&	0.143 \\
PV Cep					&311.474902&	67.960735&	102.9697&	15.2315&	341$^{6}_{-7}$&		8.228&	0.126&	-1.976&	0.110 \\\hline
\multicolumn{10}{c}{L1172/1174 complex}\\
FT Cep					&314.845315&	68.245467&	103.9926&	14.4053&	339$^{4}_{-4}$&		7.333&	0.087&	-1.599&	0.079 \\ 
2MASS J21002024+6808268	&315.084447&	68.140772&	103.9661&	14.2704&	341$^{9}_{-10}$&		7.630&	0.147&	-1.182&	0.158 \\
2MASS J21005550+6811273	&315.231481&	68.190885&	104.0418&	14.2596&	332$^{12}_{-13}$&	7.205&	0.184&	-1.649&	0.280 \\
NGC 7023 RS 2			&315.359984&	68.177338&	104.0621&	14.2140&	325$^{6}_{-6}$&		7.652&	0.104&	-1.416&	0.121 \\
NGC 7023 RS 2B			&315.362884&	68.177214&	104.0627&	14.2131&	360$^{12}_{-13}$&	6.667&	0.174&	-0.865&	0.189 \\
2MASS J21013583+6813259	&315.399160&	68.223752&	104.1087&	14.2321&	277$^{27}_{-33}$&   11.343&	0.673&	-3.039&	0.706 \\
LkH$\alpha$ 425				&315.400352&	68.139576&	104.0418&	14.1785&	332$^{6}_{-6}$&		6.971&	0.106&	-1.638&	0.117 \\
HD 200775				&315.403923&	68.163263&	104.0616&	14.1924&	357$^{5}_{-5}$&		8.336&	0.079&	-1.566&	0.083 \\
NGC 7023 RS 5			&315.427117&	68.215960&	104.1093&	14.2191&	323$^{8}_{-8}$&		7.530&	0.148&	-0.668&	0.164 \\
FU Cep					&315.444875&	68.145894&	104.0577&	14.1696&	335$^{4}_{-4}$&		7.770&	0.071&	-1.428&	0.078 \\
FV Cep					&315.558650&	68.233141&	104.1549&	14.1923&	321$^{25}_{-30}$&	8.666&	0.528&	-1.736&	0.568 \\
LkH$\alpha$ 275				&315.584980&	68.423341&	104.3129&	14.3056&	277$^{19}_{-22}$&	7.262&	0.483&	-3.589&	0.530 \\
LkH$\alpha$ 428 N				&315.617758&	68.058287&	104.0301&	14.0642&	343$^{6}_{-6}$&		6.868&	0.098&	-1.319&	0.107 \\
LkH$\alpha$ 428 S				&315.618244&	68.057669&	104.0297&	14.0637&	376$^{15}_{-16}$&	8.041&	0.193&	-1.202&	0.215 \\
FW Cep					&315.637634&	68.124746&	104.0878&	14.1008&	336$^{2}_{-2}$&		7.054&	0.044&	-1.017&	0.043 \\
NGC 7023 RS 10			&315.747855&	68.108939&	104.1022&	14.0590&	323$^{4}_{-4}$&		7.174&	0.079&	-1.738&	0.075 \\
2MASS J21025963+6808119	&315.748555&	68.136623&	104.1244&	14.0765&	401$^{85}_{-148}$&	5.678&	0.849& -11.026&	1.196 \\
EH Cep					&315.851719&	67.985134&	104.0295&	13.9501&	326$^{7}_{-7}$&		7.269&	0.140&	-2.330&	0.128 \\
2MASS J21034154+6823456	&315.923249&	68.396018&	104.3729&	14.1922&	559$^{127}_{-230}$&	5.407&	0.780&	-3.269&	0.752 \\
2MASS J21035938+6749296	&315.997585&	67.824847&	103.9389&	13.8053&	346$^{8}_{-8}$&		7.722&	0.152&	-2.061&	0.133 \\\hline
\end{tabular}
\renewcommand{\arraystretch}{1}
\end{small}
\end{table*}
%**************************************************************************************************
\bibliographystyle{aa}
\bibliography{reference}
\label{lastpage}
\end{document}